\documentclass[onecolumn,10pt,aps,pre,superscriptaddress,a4paper]{revtex4-2}

\usepackage[T1]{fontenc}
\usepackage[utf8]{inputenc}
\usepackage{amsmath}
\usepackage{amssymb}
\usepackage{bbold}
\usepackage{mathtools}
\usepackage{colonequals}
\usepackage{enumitem}
\usepackage{hyperref}

\newcommand{\bfhat}[1]{\mathbf{\hat #1}}
\newcommand{\bfhatb}[1]{\mathbf{\hat{\bar{#1}}}}
\newcommand{\mbf}[1]{\mathbf #1}
\newcommand{\mbfb}[1]{\bar{\mathbf #1}}
\newcommand{\bsy}[1]{\boldsymbol #1}

\DeclarePairedDelimiter\floor{\lfloor}{\rfloor}
\begin{document}


\title{Analytical expressions for the two basic modes of surface displacement\\ and overall deformation of a free-standing or elastically embedded sphere}

\author{Lukas Fischer}
\email{lukas.fischer@ovgu.de}
\affiliation{Institut f{\"u}r Physik, 
	Otto-von-Guericke-Universität Magdeburg, Universitätsplatz 2, 39106 Magdeburg, Germany}

\author{Andreas M. Menzel}
\email{a.menzel@ovgu.de}
\affiliation{Institut f{\"u}r Physik, 
	Otto-von-Guericke-Universität Magdeburg, Universitätsplatz 2, 39106 Magdeburg, Germany}

\date{\today}

\begin{abstract}
	Calculating by analytical theory the deformation of finite-sized elastic bodies in response to internally applied forces is a challenge. Here, we derive explicit analytical expressions for the amplitudes of modes of surface deformation of a homogeneous, isotropic, linearly elastic sphere. The modes under consideration correspond to overall changes in volume and to relative uniaxial elongation or contraction. Both situations of an elastic sphere embedded under no-slip conditions in a linearly elastic, homogeneous, isotropic, infinitely extended background medium and of a free-standing elastic sphere are considered. For example, our analytical expressions are important as a basis for computational material optimization in the context of spherical soft actuators. 
\end{abstract}

\maketitle

\section{Introduction}
\label{sec:Intro}
Studying problems of elasticity for systems that are of finite size by analytical tools of calculation is a complex task, even in the linear regime. In three-dimensional bulk, a relatively simple expression is available for the Green's function in the case of linearly elastic, homogeneous, isotropic, infinitely extended systems \cite{landau1986theory}. It quantifies the reversible displacements occurring in the system in response to a net force applied at a certain position. In that sense, the problem of linear elasticity in such three-dimensional bulk situations is --- in principle --- solved. 

Analytical investigations that include boundaries mostly concentrate on infinitely filled three-dimensional half-spaces that feature one no-slip, free-slip, or stress-free boundary. Associated Green's functions are available in the literature as well \cite{mindlin1936force, blake1971note, menzel2017force}. In that sense, these problems are --- in principle --- solved as well. More recent investigations in the context of such systems of finite extension in the direction normal to the boundary. An extreme case of this kind are strictly two-dimensional, flat, linearly elastic sheets. The associated Green's function \cite{phan1983image} features a peculiarity. It involves a logarithmic divergence with the distance from the point of application of the force. Lately, it was demonstrated that the divergence expresses an instability of the system under a net force. It cancels, in line with Newton's third law, if there is no net force acting on the elastic sheet \cite{richter2022mediated, lutz2022effect}. Recently, the case of finite, nonvanishing thickness has been addressed as well \cite{lutz2024internal}. As it turns out, the divergences found in strictly two-dimensional systems are found in that situation as well, as long as a net force is applied and the thickness remains finite. 

Still, all the described geometries are of infinite extension, even if not in all directions. Restricting the extension further to completely finite size comes at the cost of increased analytical complexity. The situation of highest symmetry would be that of a free-standing, linearly elastic sphere. It is the geometry considered in the present work. 

We build on a previous solution by Walpole \cite{walpole2002elastic}. He did not study a free-standing sphere. Instead, the linearly elastic sphere was embedded under no-slip conditions in an infinitely extended, linearly elastic, surrounding elastic background material. This includes that the traction vector (normal vector multiplied by the stress tensor) is continuous at the boundary between the elastic sphere and the elastic background. For this case, he derived the corresponding Green's function. 

One might conjecture that the situation of a free-standing elastic sphere is simply found by setting the elastic modulus of the supporting background to zero. Yet, divergences then emerge, as already noted in the original work \cite{walpole2002elastic}. In our previous considerations, we could demonstrate that these divergences are indeed and again related to the application of net forces (and/or torques) to the free-standing elastic sphere \cite{fischer2019magnetostriction}, similarly to the case of thin elastic membranes noted above. The divergences mutually cancel each other when Newton's third law is applied and no net forces and/or torques act on the system \cite{fischer2019magnetostriction}. In this case, continuity of the traction vectors means that they vanish at the surface of the elastic sphere.

The benefit of deriving corresponding expressions for the Green's function is that, analytically and exactly, one  can at any position within the sphere and on its surface calculate the reversible displacements upon elastic deformation in response to forces applied locally within the sphere. Consequently, in a type of scale-bridging approach, the overall deformation can be derived as a function of the complete discrete microscopic force distribution within the system \cite{fischer2019magnetostriction}.

To quantify and illustrate the overall deformation, an expansion of the surface displacement field into spherical harmonics is reasonable \cite{fischer2019magnetostriction}. In that context, the coefficient of the lowest mode of the displacements perpendicular to the surface, here called $u^{\bot}_{00}$, represents the amount of overall increase or decrease in volume of the sphere for positive and negative sign, respectively. Moreover, the coefficient of the second spherical harmonic associated with complete azimuthal symmetry, here termed $u^{\bot}_{20}$, quantifies uniaxial extension or contraction of the sphere along the set symmetry axis relative to the lateral directions for positive and negative sign, respectively. In previous considerations, these coefficients were determined following a relatively complicated procedure. The displacement fields were evaluated on many surface points distributed across the surface of the sphere. Then, a fit to these displacements was performed \cite{fischer2019magnetostriction}. 

In the present work, we demonstrate that this complicated procedure can be simplified substantially. Our goal is to derive analytical expressions that directly determine the coefficients $u^{\bot}_{00}$ and $u^{\bot}_{20}$ from the forces applied at a set of positions within the sphere. In fact, our analytical calculation across the following sections leads to such precise expressions. We perform the calculation using the original Green's function derived by Walpole \cite{walpole2002elastic} so that they equally apply to the case of the elastic sphere embedded under no-slip conditions in a surrounding, infinitely extended, linearly elastic background medium with continuity of the traction vectors. However, we have also confirmed that taking, at the end, the limit of vanishing elastic modulus for this background medium leads to the correct expressions for the free-standing sphere. They represent our final result. \\
The rest of the manuscript is organized as follows. In Secs.~\ref{sec:Walpole}--\ref{sec:Sol-outside-sphere}, we reproduce Walpole's solution as required for the following evaluations. In the subsequent Secs.~\ref{sec:Coordinates} and \ref{sec:Properties}, we list some necessary prerequisites. The centerpiece of our work is the calculation of the coefficient $u^{\bot}_{00}$ in Sec.~\ref{subsec:u00} and $u^{\bot}_{20}$ in Sec.~\ref{subsec:u20}. Finally, we conclude in Sec.~\ref{sec:Conclusions}.

\section{Walpole's solution}
\label{sec:Walpole}
Walpole \cite{walpole2002elastic} considers the situation when a force $\mbf{F}$ is applied within a linear elastic sphere of radius $R$ and with shear modulus $\mu$ and Poisson ratio $\nu$. Outside the elastic sphere, another, possibly different, linear elastic matrix extends to infinity, its shear modulus $\tilde{\mu}$ and Poisson ratio $\tilde{\nu}$ are denoted by a tilde. Walpole's solution allows to calculate the displacement field $\mbf{u}(\mbf{r})$ at any point, inside or outside the elastic sphere, where the center of the elastic sphere is chosen as the coordinate origin.
As we are in the linear elastic regime, the form of Walpole's solution for the displacement field $\mbf{u}(\mbf{r})$ outside the sphere is given by a matrix multiplication with the applied force $\mbf{F}$:
\begin{align}
	u_{i}(\mbf{r}) &= \sum_{j=1}^{3} \tilde{g}_{ij}(\mbf{r}, \mbfb{r}) F_j
\end{align}
Here, $ i $ identifies one coordinate of the displacement field, $ i=1,2,3 $ (or $x$, $y$, and $z$, respectively). The vectors $\mbf{r}$ and $\mbfb{r}$ denote the position of evaluation and the position where the force is applied, respectively. 
In the following, we evaluate the displacement on the surface of the elastic sphere, i.e. $ \left| \mbf{r} \right| = R$. On the surface, the inside and outside solutions produce identical results (we have also confirmed this fact numerically), as required by the no-slip conditions. 
However, the form of the outside solution is simpler. Therefore, we decide to use it in the following evaluations.

\section{Legendre polynomials}
\label{sec:Legendre}
In his solution, Walpole makes heavy use of the Legendre polynomials and its derivatives. In this work we use the following (recursive) formula \cite[Eqs.~(2.18)--(2.19)]{walpole2002elastic}:
\begin{alignat}{6}
	P_n(x)&= x \, \frac{2n-1}{n}\, P_{n-1}(x) - \frac{n-1}{n}\, P_{n-2}(x), \qquad
	&&P_0(x)&&=1,\qquad
	&&P_1(x)&&=x,\\
	P'_{n+1}(x)&=(2n+1)P_n(x)+P'_{n-1}(x),\qquad
	&&P'_{1}(x)&&=1,\qquad
	&&P'_{2}(x)&&=3x,\\
	P''_{n+2}(x)&=(2n+3)P'_{n+1}(x)+P''_{n}(x),\qquad
	&&P''_{2}(x)&&=3,\qquad
	&&P''_{3}(x)&&=15x.
\end{alignat}
We now define (following Walpole's notation, but omitting the normalization factor) for $ n \in \mathbb{N}_0 $ ($ n $ is a superscript, not an exponent):
\begin{align}
	H^n &= P_n(x), \\
	K^n &= P'_{n+1}(x), \\
	L^n &= P''_{n+2}(x).
\end{align}
\newpage\noindent
The parameter $ x $ is defined by (for $ \mbf{r},\, \mbfb{r} \neq \bsy{0} $)
\begin{align}
	\label{x-eq}
	x = \frac{\mbf{r}\cdot\mbfb{r}}{r \bar{r}}, 
\end{align}
so it depends on the position of evaluation and the position of the force center. Here, we introduced the notation $r=|\mbf{r}|$ and $\bar{r}=|\mbfb{r}|$. This definition was also employed by Walpole. \\
The dependence on $ x $ is from now on implicitly assumed for $ H^n $, $ K^n $, and $ L^n $.
\section{Constants}
\label{sec:Constants}
We use the following constants, all depending only on $ \mu $, $ \tilde{\mu} $, $ \nu $, $ \tilde{\nu} $ and index $ n \in \mathbb{N}_0 $ (and not on the positions of the force or the position of evaluation):
\begin{align}
	\tilde{q}_n &= \left[ n^2+(1-2\tilde{\nu})n+1-\tilde{\nu}\right] \tilde{\mu} + (n-1)\left[ (3-4\tilde{\nu})n+2(1-\tilde{\nu}) \right] \mu, \label{eq:constants_1}\\
	q_n &= (n+2)\left[ (3-4\nu)n+(1-2\nu) \right]\tilde{\mu} + \left[ (n-1)^2 +  (3+2\nu)(n-1)+3(1+\nu) \right] \mu, \\
	a_n &= \frac{n+2}{(n-1)\mu+(n+2)\tilde{\mu}},\\
	b_n &= \frac{2\mu}{\tilde{q}_{n+1}} \, \frac{\left[ (1-2\nu)n+3(1-\nu) \right] \left[ (3-4\tilde{\nu})n+ 5-6\tilde{\nu} \right]}{(n-1)\mu+(n+2)\tilde{\mu}},\\
	c_n &= \frac{2\mu}{\tilde{q}_{n+1}} \, \frac{2n+1}{(n-1)\mu+(n+2)\tilde{\mu}}\\
	d_n &= \epsilon_n \frac{\mu}{q_{n-1}} \, \frac{(1-2\nu)n-2+\nu}{(n-1)\mu+(n+2)\tilde{\mu}} \quad\text{ with } 
	\epsilon_n=\begin{cases} 1, n>0 \\ 0, n=0 \end{cases}.
\end{align}
Let us note that we suppose here $ \mu \ne 2\tilde{\mu} $ in order to avoid a divergence in the above expressions.
\begin{align}
	e_n &= \mu(1-\nu)(2n+1)/\tilde{q}_{n+1}, \\
	m_n &= \mu (1-\nu)(2n+1)(2n-1)/q_{n-1} \text{ for } n \geq 1,\\
	s_n &= a_n+ (2n+1)\left[ n+8(1-\nu) \right]d_n, \\
	h_n &= \frac{1+(\mu-\tilde{\mu})(a_n-b_n)}{2n+3} + (\nu-\tilde{\nu})\tilde{\mu}c_n + (\mu-\tilde{\mu})\left[ n+4(1-\nu)) \right]d_n,\\
	p_n &= (3-4\nu)a_n-(2n+1)\left[ n+4(1-\nu) \right]d_n, \\
	k_n &= 3-4\nu + (\mu-\tilde{\mu}) p_n, \\
	l_n &= 1+(\mu-\tilde{\mu})s_n, \\
	u_n &= \mu (1-\nu) \left[\frac{2n+1}{q_{n-1}} -\frac{2n-3}{\tilde{q}_{n-1}}\right]. \label{eq:constants_14}
\end{align}
\newpage

\section{Solution outside the sphere}
\label{sec:Sol-outside-sphere}
The solution for the matrix $ \tilde{g}_{ij}(\mbf{r}) $ as given by Walpole reads:
\begin{align}
	\label{eq:sol}
	16\pi\mu(1-\nu)\tilde{g}_{ij}(\mbf{r}) &= \tilde{G}_1 r^2 \delta_{ij} + \tilde{G}_2  r_i  r_j + \tilde{G}_3 \bar{r}_i \bar{r}_j
	+ \tilde{G}_4( r_i\bar{r}_j-\bar{r}_i r_j) + \tilde{G}_5( r_i\bar{r}_j+\bar{r}_i r_j) \\
	\text{with } \tilde{G}_N &= \frac{1}{r^5} \sum_{n=0}^{\infty} \left(\frac{\bar{r}}{r}\right)^n \left[ r^2 S_N^n + R^2 T_N^n \right], 
	\qquad N \in \{1,2,3,4,5\}. \label{eq:sol_Walpole}
\end{align}
Here, we use the following expressions depending on the constants and Legendre polynomials:
\begin{align}
	S_1^n &= (h_n-e_n)K^n + k_n H^n, \\
	S_2^n &= (2n+3)h_n K^n - (h_n-e_n)L^n, \\
	S_3^n &= m_{n+2} K^n - (h_{n+2}-e_{n+2})L^n, \\
	S_4^n &= \frac{1}{2} \left[ l_{n+1}-(2n+5)h_{n+1} \right]K^n, \\
	S_5^n &= (h_{n+1}-e_{n+1})L^n - \frac{1}{2} \left[ (2n+5)h_{n+1} + m_{n+1} \right]K^n, \\
	T_1^n &= -u_{n+2}K^n, \\
	T_2^n &= u_{n+2}L^n, \\
	T_3^n &= u_{n+4}L^n, \\
	T_4^n &= 0, \\
	T_5^n &= -u_{n+3}L^n.
\end{align}
The prefactor $ \left(\frac{\bar{r}}{r}\right)^n $ guarantees convergence of the series because outside the sphere $ r > \bar{r} $ and the other terms in the product grow at most linearly with $ n $. It is known that the series over $ n q^n $ converges for $ |q|<1 $.\\
For the following calculation, we note that $r^2 = R^2$ on the surface of the elastic sphere, so both $S_N^n$ and $T_N^n$ contribute equally in Eq.~\eqref{eq:sol_Walpole}.

\newpage
\section{Coordinates and spherical harmonics}
\label{sec:Coordinates}
From this section forward, our own calculation starts.\\
In general, we can express the spherical coordinates of $ \bfhat{r} = \mbf{r} / r $ using spherical harmonics:
\begin{alignat}{3}
	\left(\bfhat{r}\right)_x &= \sin\theta \, \cos\varphi &&= - \sqrt{\frac{2\pi}{3}} \left( Y_{11}(\theta, \varphi) - Y_{1-1}(\theta, \varphi)\right), \label{eq:r_x}\\
	\left(\bfhat{r}\right)_y &= \sin\theta \, \sin\varphi &&= i \sqrt{\frac{2\pi}{3}} \left( Y_{11}(\theta, \varphi) + Y_{1-1}(\theta, \varphi)\right), \label{eq:r_y}\\
	\left(\bfhat{r}\right)_z &= \cos\theta &&= \sqrt{\frac{4\pi}{3}} Y_{10}(\theta, \varphi). \label{eq:r_z}
\end{alignat}
Here and also in the following, $ i $ denotes the imaginary unit. \\
Additionally, the dyadic product $ \bfhat{r}\bfhat{r} $ can also be expressed in terms of spherical harmonics:
\begin{alignat}{3}
	\left(\bfhat{r}\bfhat{r}\right)_{xx} &= \sin^2\theta \, \cos^2\varphi &&=
	\sqrt{\frac{2\pi}{15}} \left( Y_{22}(\theta, \varphi) + Y_{2-2}(\theta, \varphi)\right) - \sqrt{\frac{4\pi}{9}} \left( -Y_{00}(\theta, \varphi) + \sqrt{\frac{1}{5}} Y_{20}(\theta, \varphi)\right), \label{eq:rr_xx}\\[10pt]
	\left(\bfhat{r}\bfhat{r}\right)_{xy} &= \sin^2\theta \, \cos\varphi \, \sin\varphi &&= -i \sqrt{\frac{2\pi}{15}} \left( Y_{22}(\theta, \varphi) - Y_{2-2}(\theta, \varphi)\right), \label{eq:rr_xy}\\[10pt]
	\left(\bfhat{r}\bfhat{r}\right)_{yy} &= \sin^2\theta \, \sin^2\varphi &&= -\sqrt{\frac{2\pi}{15}} \left( Y_{22}(\theta, \varphi) + Y_{2-2}(\theta, \varphi)\right) - \sqrt{\frac{4\pi}{9}} \left( -Y_{00}(\theta, \varphi) + \sqrt{\frac{1}{5}} Y_{20}(\theta, \varphi)\right), \label{eq:rr_yy}\\[10pt]
	\left(\bfhat{r}\bfhat{r}\right)_{xz} &= \sin\theta \, \cos\theta \, \cos\varphi &&= - \sqrt{\frac{2\pi}{15}} \left(Y_{21}(\theta, \varphi) - Y_{2-1}(\theta, \varphi)\right), \label{eq:rr_xz}\\
	\left(\bfhat{r}\bfhat{r}\right)_{yz} &= \sin\theta \, \cos\theta \, \sin\varphi &&= i \sqrt{\frac{2\pi}{15}} \left(Y_{21}(\theta, \varphi) + Y_{2-1}(\theta, \varphi)\right), \label{eq:rr_yz}\\
	\left(\bfhat{r}\bfhat{r}\right)_{zz} &= \cos^2\theta &&= \sqrt{\frac{4\pi}{9}} \left(Y_{00}(\theta, \varphi) + \sqrt{\frac{4}{5}} Y_{20}(\theta, \varphi)\right). \label{eq:rr_zz}
\end{alignat}

Obviously, this dyadic product is symmetric, so that we have $ \left(\bfhat{r}\bfhat{r}\right)_{ij} = \left(\bfhat{r}\bfhat{r}\right)_{ji} $ and the components listed above already cover all possible index combinations.

\newpage
\section[Properties]{Useful properties of the Legendre polynomials and relations to the spherical harmonics}
\label{sec:Properties}

Walpole's solution contains derivatives of the Legendre polynomials. They can be expressed again in terms of Legendre polynomials as follows \cite{whittaker1920course}:
\begin{align}
	P'_{n+1}(x) &= \sum_{k=0}^{\floor*{\frac{n}{2}}} \left(2(n-2k)+1\right) P_{n-2k}(x), \label{eq:deriv1}\\
	P''_{n+2}(x) &= \sum_{k=0}^{\floor*{\frac{n}{2}}} \left(2(n-2k)+3\right) \sum_{l=0}^{\floor*{\frac{n}{2}}-k} \left(2(n-2(k+l))+1\right) P_{n-2(k+l)}(x). \label{eq:deriv2}
\end{align}
Here, $\floor*{\frac{n}{2}}$ denotes the floor function, so $\floor*{\frac{n}{2}} = \frac{n}{2}$ for $n$ even and $\frac{n-1}{2}$ for $n$ odd.
The second equation follows from the first by differentiating both sides and using the formula again for the derivative on the right-hand side.\\
Furthermore, we can express the Legendre polynomials at $ x = \bfhat{r} \cdot \bfhatb{r} $ in terms of spherical harmonics at positions $ \bfhat{r}(\theta, \varphi) $ and $ \bfhatb{r}(\bar{\theta}, \bar{\varphi}) $ \cite{whittaker1920course}:
\begin{align}
	\label{eq:add_sph}
	P_l(x) &= \frac{4\pi}{2l+1} \sum_{m=-l}^{l} Y_{lm}^*(\theta, \varphi) Y_{lm}(\bar{\theta}, \bar{\varphi}),
\end{align}
where the star denotes complex conjugation. \\
In the following, we use the shorthand notation
\begin{align}
	Y_{lm} &\coloneqq Y_{lm}(\theta, \varphi), \\
	\bar{Y}_{lm} &\coloneqq Y_{lm}(\bar{\theta}, \bar{\varphi}). \label{eq:sph_harmonics_abbrv}
\end{align}
We can also express products of spherical harmonics in terms of a sum of single spherical harmonics \cite{sakuraiqm}:
\begin{align}
	\label{eq:prod_sph}
	Y_{a,\alpha}(\theta, \varphi) Y_{b,\beta}(\theta, \varphi) &=\sqrt{\frac{\left(2a+1\right)\left(2b+1\right)}{4\pi}}\sum_{c,\gamma}\left(-1\right)^{\gamma}\sqrt{2c+1}\left(\begin{array}{ccc}
		a & b & c\\
		\alpha & \beta & -\gamma
	\end{array}\right)\left(\begin{array}{ccc}
		a & b & c\\
		0 & 0 & 0
	\end{array}\right) Y_{c,\gamma}(\theta, \varphi),
\end{align}
where $ \left(\begin{array}{ccc}
	a & b & c\\
	\alpha & \beta & -\gamma
\end{array}\right) $ refers to the Wigner $ 3j $-symbols. \newpage
Application of Eq.~\eqref{eq:prod_sph} leads to the following identities:
\begin{align}
	Y_{10}Y_{20} &= \sqrt{\frac{15}{4\pi}} \left(\frac{2\sqrt{3}}{15}\, Y_{10} + \frac{3\sqrt{7}}{35}\, Y_{30} \right), \label{eq:identities_1}\\
	Y_{11}Y_{20} &= \sqrt{\frac{1}{20\pi}} \left(- Y_{11} + \sqrt{\frac{18}{7}}\, Y_{31} \right), \label{eq:identities_2}\\
	Y_{1-1}Y_{20} &= \sqrt{\frac{1}{20\pi}} \left(- Y_{1-1} + \sqrt{\frac{18}{7}}\, Y_{3-1} \right), \label{eq:identities_3}\\
	Y_{20}Y_{20} &= \sqrt{\frac{1}{4\pi}} \left(Y_{00} + \frac{2\sqrt{5}}{7}\, Y_{20} +\frac{6}{7}\, Y_{40} \right), \label{eq:identities_4}\\
	Y_{21}Y_{20} &= \frac{1}{7}\sqrt{\frac{5}{4\pi}} \left(Y_{21} + \sqrt{6}\, Y_{41} \right), \label{eq:identities_5}\\
	Y_{2-1}Y_{20} &= \frac{1}{7}\sqrt{\frac{5}{4\pi}} \left(Y_{2-1} + \sqrt{6}\, Y_{4-1} \right), \label{eq:identities_6}\\
	Y_{22}Y_{20} &= \frac{1}{7}\sqrt{\frac{5}{4\pi}} \left(-2\, Y_{22} + \sqrt{3}\, Y_{42} \right), \label{eq:identities_7}\\
	Y_{2-2}Y_{20} &= \frac{1}{7}\sqrt{\frac{5}{4\pi}} \left(-2\, Y_{2-2} + \sqrt{3}\, Y_{4-2} \right). \label{eq:identities_8}
\end{align}
\newpage
\section{Walpole's solution projected onto a spherical harmonic}
\label{sec:Calculation}
\subsection{Calculation of $u^{\bot}_{00}$}
\label{subsec:u00}
Our aim is to evaluate the following mode of deformation:
\begin{align}
	u^{\bot}_{00} = \int_{0}^{\pi} d\theta\, \sin\theta \int_{0}^{2\pi}d\varphi \; \mbf{u}(\mbf{r}) \cdot \bfhat{r} \; Y_{00}.
\end{align}
Here, we need the simple identity 
\begin{align}
	Y_{00} Y_{lm} &= \sqrt{\frac{1}{4\pi}}\, Y_{lm} \label{eq:Y_00}
\end{align}
which immediately follows from the definition of $Y_{00}$.\\
For the following calculation, we consider each contribution from Eq.~\eqref{eq:sol} separately.
\begin{enumerate}[label=\Roman*.]
	
	\item $ \tilde{G}_1 r^2 \mbf{F} $. \\
	Projecting this term onto $ \bfhat{r} $ on the spherical surface of radius $ R $ (as required for the $\bot$-component of $\mbf{u}$) and using Eqs.~\eqref{eq:r_x}--\eqref{eq:r_z} as well as Eq.~\eqref{eq:Y_00} leads to:
	\begin{align}
		\mbf{F} \cdot \bfhat{r}\; Y_{00} &= \sqrt{\frac{1}{6}}\left\{ -F_x \left( Y_{11} - Y_{1-1}\right) +i\, F_y \left( Y_{11} + Y_{1-1}\right) +F_z \sqrt{2}\, Y_{10} \right\}. 
		\label{eq:G1_sph2}
	\end{align}
	\begin{enumerate}[label=\alph*.]
		\item $ k_n H^n $.
		Using the orthonormalization of the spherical harmonics and Eq.~\eqref{eq:add_sph} for $H^n$ [notice the shorthand notation from Eq.~\eqref{eq:sph_harmonics_abbrv}], we calculate:
		\begin{align}
			&\int_{0}^{\pi} d\theta\, \sin\theta \int_{0}^{2\pi}d\varphi \; k_n H^n R^2 \mbf{F} \cdot \bfhat{r}\; Y_{00} \nonumber \\
			&\quad = \begin{cases}
				\frac{4\pi\, k_n R^2}{3 \sqrt{6}}\left\{ -F_x \left( \bar{Y}_{11} - \bar{Y}_{1-1}\right) +i\, F_y \left( \bar{Y}_{11} + \bar{Y}_{1-1}\right) +F_z \sqrt{2}\, \bar{Y}_{10} \right\}, \quad n=1, \\
				0, \quad \text{else.}
			\end{cases}
			\label{eq:G1a2}
		\end{align}
		\item $ (h_n-e_n-u_{n+2}) K^n = \alpha_n K^n $.\\
		We introduced the definition $ \alpha_n \coloneqq h_n-e_n-u_{n+2} $. Here and in the following, we denote by Greek letters the constants that we introduce ourselves (those that are not defined in the original Walpole paper).\\
		By first expanding the derivative using Eq.~\eqref{eq:deriv1}, then again using the orthonormalization of the spherical harmonics and Eq.~\eqref{eq:add_sph}, we get:
		\begin{align}
			&\int_{0}^{\pi} d\theta\, \sin\theta \int_{0}^{2\pi}d\varphi \; \alpha_n K^n R^2 \mbf{F} \cdot \bfhat{r}\; Y_{00} \nonumber \\
			&\quad = \begin{cases}
				\frac{4\pi\, \alpha_n R^2}{\sqrt{6}}\left\{ -F_x \left( \bar{Y}_{11} - \bar{Y}_{1-1}\right) +i\, F_y \left( \bar{Y}_{11} + \bar{Y}_{1-1}\right) +F_z \sqrt{2}\, \bar{Y}_{10} \right\}, \quad n \geq 1 \text{ odd}, \\
				0, \quad \text{else.}
			\end{cases} 
			\label{eq:G1b2}
		\end{align}
	\end{enumerate}
	
	\item $ \tilde{G}_2 (\mbf{r}\mbf{r}) \cdot \mbf{F} $. \\
	Projecting onto $ \bfhat{r} $ on the spherical surface of radius $ R $ leads to:
	\begin{align}
		\tilde{G}_2 \left((\mbf{r}\mbf{r}) \cdot \mbf{F} \right) \cdot \bfhat{r} &= \tilde{G}_2 R^2 \bfhat{r} \cdot \mbf{F}.
	\end{align}
	From here, we proceed as in Eq.~\eqref{eq:G1_sph2}.
	\begin{enumerate}[label=\alph*.]
		\item $ (2n+3)h_n K^n $.
		The result is the same as in Eq.~\eqref{eq:G1b2}, replacing $ \alpha_n $ by $ (2n+3)h_n $.
		\item $ -\alpha_n L^n $. Here, we use Eq.~\eqref{eq:deriv2} and proceed as before:
		\begin{align}
			&-\int_{0}^{\pi} d\theta\, \sin\theta \int_{0}^{2\pi}d\varphi \; \alpha_n L^n R^2 \mbf{F} \cdot \bfhat{r}\; Y_{00} \nonumber \\
			&\quad = \begin{cases}
				-\frac{4\pi\, \alpha_n R^2}{\sqrt{6}}\Big\{ -F_x \left( \bar{Y}_{11} - \bar{Y}_{1-1}\right) +i\, F_y \left( \bar{Y}_{11} + \bar{Y}_{1-1}\right) +F_z \sqrt{2}\, \bar{Y}_{10} \Big\} \frac{n+1}{2}(n+4), \quad n \geq 1 \text{ odd}, \\
				0, \quad \text{else.}
			\end{cases}
			\label{eq:G2b2}
		\end{align}
	\end{enumerate}
	\item $ \tilde{G}_3 (\mbfb{r}\mbfb{r}) \cdot \mbf{F} $. \\
	Projecting onto $ \bfhat{r} $ on the spherical surface of radius $ R $ leads to:
	\begin{align}
		\tilde{G}_3 \left((\mbfb{r}\mbfb{r}) \cdot \mbf{F} \right) \cdot \bfhat{r} &= \tilde{G}_3\, \mbfb{r} \cdot \mbf{F} \; \mbfb{r} \cdot \bfhat{r}.
	\end{align}
	From here, we proceed in analogy to Eq.~\eqref{eq:G1_sph2}.
	\begin{align}
		\mbfb{r} \cdot \bfhat{r}\; Y_{00} &= \sqrt{\frac{1}{6}}\left\{ -\bar{r}_x \left( Y_{11} - Y_{1-1}\right) +i\, \bar{r}_y \left( Y_{11} + Y_{1-1}\right) +\bar{r}_z \sqrt{2}\, Y_{10} \right\}.  \label{eq:G3_sph2}
	\end{align}
	\begin{enumerate}[label=\alph*.]
		\item $ m_{n+2} K^n $.
		The result is similar to Eq.~\eqref{eq:G1b2}:
		\begin{align}
			&\int_{0}^{\pi} d\theta\, \sin\theta \int_{0}^{2\pi}d\varphi \; m_{n+2} K^n \; \mbfb{r} \cdot \mbf{F} \; \mbfb{r} \cdot \bfhat{r}\; Y_{00} \nonumber \\
			&\quad = \begin{cases}
				\mbfb{r} \cdot \mbf{F}\; \frac{4\pi\, m_{n+2}}{\sqrt{6}}\left\{ -\bar{r}_x \left( \bar{Y}_{11} - \bar{Y}_{1-1}\right) +i\, \bar{r}_y \left( \bar{Y}_{11} + \bar{Y}_{1-1}\right) +\bar{r}_z \sqrt{2}\, \bar{Y}_{10} \right\}, \quad n \geq 1 \text{ odd}, \\
				0, \quad \text{else.}
			\end{cases} \label{eq:G3a2}
		\end{align}
		\item $ -\alpha_{n+2} L^n $.
		The result is similar to Eq.~\eqref{eq:G2b2}:
		\begin{align}
			&-\int_{0}^{\pi} d\theta\, \sin\theta \int_{0}^{2\pi}d\varphi \; \alpha_{n+2} L^n \; \mbfb{r} \cdot \mbf{F} \; \mbfb{r} \cdot \bfhat{r}\; Y_{20} \nonumber \\
			&\quad = \begin{cases}
				-\mbfb{r} \cdot \mbf{F}\; \frac{4\pi\, \alpha_{n+2}}{\sqrt{6}}\Big\{ -\bar{r}_x \left( \bar{Y}_{11} - \bar{Y}_{1-1}\right) +i\, \bar{r}_y \left( \bar{Y}_{11} + \bar{Y}_{1-1}\right) \\
				\qquad +\bar{r}_z \sqrt{2}\, \bar{Y}_{10} \Big\} \frac{n+1}{2}(n+4), \quad n \geq 1 \text{ odd}, \\
				0, \quad \text{else.}
			\end{cases}
			\label{eq:G3b2}
		\end{align}
	\end{enumerate}
	\item $ \tilde{G}_4 (\mbf{r}\mbfb{r}-\mbfb{r}\mbf{r}) \cdot \mbf{F} $. \\
	Projecting onto $ \bfhat{r} $ on the spherical surface of radius $ R $ leads to [using Eqs.~\eqref{eq:rr_xx}--\eqref{eq:rr_zz}]
	\begin{align}
		\tilde{G}_4 \left( (\mbf{r}\mbfb{r}-\mbfb{r}\mbf{r}) \cdot \mbf{F} \right) \cdot \bfhat{r} &= \tilde{G}_4\,R \left( \mbfb{r} \cdot \mbf{F} - \mbf{F} \cdot \bfhat{r}\; \bfhat{r} \cdot \mbfb{r} \right). \label{eq:G4_1}
	\end{align}
	When we treat the terms from the difference in the brackets in Eq.~\eqref{eq:G4_1} separately, we get:
	\begin{align}
		\mbfb{r} \cdot \mbf{F} \; Y_{00} &= \mbfb{r} \cdot \mbf{F} \; Y_{00} \text{ (already in final form)}, \label{eq:G42_2}\\
		\mbf{F} \cdot \bfhat{r}\; \bfhat{r} \cdot \mbfb{r}\; Y_{00}
		&= \sqrt{\frac{1}{30}} \Big[ \left( Y_{22} + Y_{2-2} \right) \left(F_x \bar{r}_x - F_y \bar{r}_y \right)
		-i \left( Y_{22} - Y_{2-2} \right) \left(F_x \bar{r}_y + F_y \bar{r}_x \right) \nonumber \\
		&\qquad \qquad -\left(Y_{21} - Y_{2-1} \right) \left(F_x \bar{r}_z + F_z \bar{r}_x \right) +i \left(Y_{21} + Y_{2-1} \right) \left(F_y \bar{r}_z + F_z \bar{r}_y \right) \Big]\nonumber \\
		&\qquad \qquad + \frac{1}{3} \Bigg[ \left( Y_{00} - \sqrt{\frac{1}{5}} Y_{20} \right) \left(F_x \bar{r}_x + F_y \bar{r}_y \right)+ \left(Y_{00} + \sqrt{\frac{4}{5}} Y_{20} \right) F_z \bar{r}_z \Bigg]. \label{eq:G42_3}
	\end{align}
	$ \tilde{G}_4$ only contains one term: $ \frac{1}{2} \left[ l_{n+1}-(2n+5)h_{n+1} \right]K^n = \beta_n K^n$ \\
	where we introduced the definition $ \beta_n \coloneqq \frac{1}{2} \left[ l_{n+1}-(2n+5)h_{n+1} \right] $.
	\begin{align}
		&\int_{0}^{\pi} d\theta\, \sin\theta \int_{0}^{2\pi}d\varphi \; \beta_n K^n R\; \left(\mbfb{r} \cdot \mbf{F} - \mbf{F} \cdot \bfhat{r}\; \bfhat{r} \cdot \mbfb{r}\right) \, Y_{00} \nonumber \\
		&\quad = \beta_n R \; 4\pi \begin{cases}
			\mbfb{r} \cdot \mbf{F}\; \bar{Y}_{00} - \frac{1}{3} \bar{Y}_{00} \left[ \left(F_x \bar{r}_x + F_y \bar{r}_y \right)+ F_z \bar{r}_z \right], \quad n=0,\\
			\mbfb{r} \cdot \mbf{F}\; \bar{Y}_{00} - \frac{1}{3} \bar{Y}_{00} \left[ \left(F_x \bar{r}_x + F_y \bar{r}_y \right)+ F_z \bar{r}_z \right]\\
			\qquad - \sqrt{\frac{1}{30}} \Big[ \left( \bar{Y}_{22} + \bar{Y}_{2-2} \right) \left(F_x \bar{r}_x - F_y \bar{r}_y \right)
			-i \left( \bar{Y}_{22} - \bar{Y}_{2-2} \right) \left(F_x \bar{r}_y + F_y \bar{r}_x \right) \\
			\qquad -\left(\bar{Y}_{21} - \bar{Y}_{2-1} \right) \left(F_x \bar{r}_z + F_z \bar{r}_x \right) 
			+i \left(\bar{Y}_{21} + \bar{Y}_{2-1} \right) \left(F_y \bar{r}_z + F_z \bar{r}_y \right) \Big] \\
			\qquad -\frac{1}{3\sqrt{5}} \bar{Y}_{20} \left[ -\left(F_x \bar{r}_x + F_y \bar{r}_y \right)+2 F_z \bar{r}_z \right],	\quad n \geq 2 \text{ even},\\
			0, \quad \text{else.}
		\end{cases} \label{eq:G42}
	\end{align}
	The term for $n=0$ as well as the first line for $n=2$ can be simplified with the following equation:
	\begin{align}
		\mbfb{r} \cdot \mbf{F} -\frac{1}{3} \left[ \left(F_x \bar{r}_x + F_y \bar{r}_y \right)+ F_z \bar{r}_z \right] &= \frac{2}{3}\, \mbfb{r} \cdot \mbf{F}.
		\label{eq:G42_simpl}
	\end{align}
	\item $ \tilde{G}_5 (\mbf{r}\mbfb{r}+\mbfb{r}\mbf{r}) \cdot \mbf{F} $. \\
	We can use Eq.~\eqref{eq:G42_2} and Eq.~\eqref{eq:G42_3} again, this time summing the two resulting terms.
	\begin{enumerate}[label=\alph*.]
		\item $ - \frac{1}{2} \left[ (2n+5)h_{n+1} + m_{n+1} \right]K^n = -\gamma_n K^n$ \\
		with the definition $ \gamma_n \coloneqq \frac{1}{2} \left[ (2n+5)h_{n+1} + m_{n+1} \right] $.
		\begin{align}
			&-\int_{0}^{\pi} d\theta\, \sin\theta \int_{0}^{2\pi}d\varphi \; \gamma_n K^n R\; \left(\mbfb{r} \cdot \mbf{F} + \mbf{F} \cdot \bfhat{r}\; \bfhat{r} \cdot \mbfb{r}\right) \, Y_{00} \nonumber \\
			&\quad = -\gamma_n R \; 4\pi \begin{cases}
				\mbfb{r} \cdot \mbf{F}\; \bar{Y}_{00} + \frac{1}{3} \bar{Y}_{00} \left[ \left(F_x \bar{r}_x + F_y \bar{r}_y \right)+ F_z \bar{r}_z \right], \quad n=0,\\
				\mbfb{r} \cdot \mbf{F}\; \bar{Y}_{00} + \frac{1}{3} \bar{Y}_{00} \left[ \left(F_x \bar{r}_x + F_y \bar{r}_y \right)+ F_z \bar{r}_z \right]\\
				\qquad + \sqrt{\frac{1}{30}} \Big[ \left( \bar{Y}_{22} + \bar{Y}_{2-2} \right) \left(F_x \bar{r}_x - F_y \bar{r}_y \right) 
				-i \left( \bar{Y}_{22} - \bar{Y}_{2-2} \right) \left(F_x \bar{r}_y + F_y \bar{r}_x \right) \\
				\qquad -\left(\bar{Y}_{21} - \bar{Y}_{2-1} \right) \left(F_x \bar{r}_z + F_z \bar{r}_x \right) 
				+i \left(\bar{Y}_{21} + \bar{Y}_{2-1} \right) \left(F_y \bar{r}_z + F_z \bar{r}_y \right) \Big] \Big]\\
				\qquad +\frac{1}{3\sqrt{5}} \bar{Y}_{20} \left[ -\left(F_x \bar{r}_x + F_y \bar{r}_y \right)+2 F_z \bar{r}_z \right],	\quad n \geq 2 \text{ even},\\
				0, \quad \text{else.}
			\end{cases} \label{eq:G5a2}
		\end{align}
		Similar to Eq.~\eqref{eq:G42_simpl}, we have:
		\begin{align}
			\mbfb{r} \cdot \mbf{F} +\frac{1}{3} \left[ \left(F_x \bar{r}_x + F_y \bar{r}_y \right)+ F_z \bar{r}_z \right] &= \frac{4}{3}\, \mbfb{r} \cdot \mbf{F}
			\label{eq:G5_simpl2}
		\end{align}
		\item $ \alpha_{n+1} L^n $.
		\begin{align}
			&\int_{0}^{\pi} d\theta\, \sin\theta \int_{0}^{2\pi}d\varphi \; \alpha_{n+1} L^n R\; \left(\mbfb{r} \cdot \mbf{F} + \mbf{F} \cdot \bfhat{r}\; \bfhat{r} \cdot \mbfb{r}\right) \, Y_{00} \nonumber \\
			&\quad = \alpha_{n+1} R \; 4\pi \begin{cases}
				4 \mbfb{r} \cdot \mbf{F}\; \bar{Y}_{00}, \quad n=0,\\
				2(n+2)\frac{n+3}{3} \mbfb{r} \cdot \mbf{F}\; \bar{Y}_{00}\\
				\qquad + \frac{n}{2} (n+5)\sqrt{\frac{1}{30}} \Big[ \left( \bar{Y}_{22} + \bar{Y}_{2-2} \right) \left(F_x \bar{r}_x - F_y \bar{r}_y \right) \\
				\qquad -i \left( \bar{Y}_{22} - \bar{Y}_{2-2} \right) \left(F_x \bar{r}_y + F_y \bar{r}_x \right) \\
				\qquad -\left(\bar{Y}_{21} - \bar{Y}_{2-1} \right) \left(F_x \bar{r}_z + F_z \bar{r}_x \right) \\
				\qquad +i \left(\bar{Y}_{21} + \bar{Y}_{2-1} \right) \left(F_y \bar{r}_z + F_z \bar{r}_y \right) \Big] \\
				\qquad +\frac{n}{2} \frac{n+5}{3\sqrt{5}} \bar{Y}_{20} \left[ -\left(F_x \bar{r}_x + F_y \bar{r}_y \right)+2 F_z \bar{r}_z \right], \quad n \geq 2 \text{ even}, \\
				0, \quad \text{else.}
			\end{cases} \label{eq:G5b2}
		\end{align}
	\end{enumerate}
	
\end{enumerate}
In the last equation, the case $ n \geq 2 $ is also valid for $ n=0 $ because $ \frac{n}{2}=0 $ in that case, so the other terms in the square brackets do not contribute.\\

\noindent
The terms I--V encompass all contributions to $ u^{\bot}_{00} $. Thus, we can get the final result for $ u^{\bot}_{00} $ by summing up all contributions from I--V and by adding terms for all $ n \in \mathbb{N}_0 $ (with a prefactor of $ \left(\frac{\bar{r}}{r}\right)^n $). Furthermore, the whole result needs to be divided by $ 16\pi\mu(1-\nu)$ and $R^3$ [from the prefactor of the sum $\frac{1}{R^5}$ and $R^2$ from the prefactor of $S_N^n$ and $T_N^n$ in Eq.~\eqref{eq:sol_Walpole}]. \newpage

Next, we simplify these terms by applying Eqs.~\eqref{eq:r_x}--\eqref{eq:r_z} in reverse, but for $\mbfb{r}$ (replacing $\bfhat{r}$ by $\bfhatb{r}$ and $Y_{lm}$ by $\bar{Y}_{lm}$) to express the spherical harmonics in terms of the coordinates of the position $\mbfb{r}$ where the force is applied. Obviously, we can use $\bar{Y}_{00} = Y_{00}$ since this expression is a constant. Therefore, the terms I--V from above are rewritten as follows:
\begin{enumerate}[label=\Roman*.]
	\item We rewrite the curly brackets (for a and b):
	\begin{align}
		& \frac{1}{\sqrt{6}}\left\{ -F_x \left( \bar{Y}_{11} - \bar{Y}_{1-1}\right) +i\, F_y \left( \bar{Y}_{11} + \bar{Y}_{1-1}\right) +F_z \sqrt{2}\, \bar{Y}_{10} \right\} \nonumber\\
		= & Y_{00} \left\{ F_x \bfhatb{r}_x + F_y \bfhatb{r}_y + F_z \bfhatb{r}_z\right\} = Y_{00} \; \mbf{F} \cdot \bfhatb{r}. \label{eq:G1_ab2}
	\end{align}
	\item  Same expression in the curly brackets.
	\item Similar expression in the curly brackets as in Eq.~\ref{eq:G1_ab2}:
	\begin{align}
		& \frac{1}{\sqrt{6}}\left\{ -\bar{r}_x \left( \bar{Y}_{11} - \bar{Y}_{1-1}\right) +i\, \bar{r}_y \left( \bar{Y}_{11} + \bar{Y}_{1-1}\right) +\bar{r}_z \sqrt{2}\, \bar{Y}_{10} \right\} \nonumber\\
		= & Y_{00} \; \mbfb{r} \cdot \bfhatb{r} = Y_{00} \bar{r}. \label{eq:G3_ab2}
	\end{align}
	\item Expression for $n \geq 2$ (without first term and negative sign) by rearranging the terms (first equality) and then using Eqs.~\eqref{eq:rr_xx}--\eqref{eq:rr_zz} (second equality):
	{\allowdisplaybreaks
	\begin{align}
		&\sqrt{\frac{1}{30}} \Big[ \left( \bar{Y}_{22} + \bar{Y}_{2-2} \right) \left(F_x \bar{r}_x - F_y \bar{r}_y \right) -i \left( \bar{Y}_{22} - \bar{Y}_{2-2} \right) \left(F_x \bar{r}_y + F_y \bar{r}_x \right) \nonumber\\
		& -\left(\bar{Y}_{21} - \bar{Y}_{2-1} \right) \left(F_x \bar{r}_z + F_z \bar{r}_x \right) +i \left(\bar{Y}_{21} + \bar{Y}_{2-1} \right) \left(F_y \bar{r}_z + F_z \bar{r}_y \right) \Big] \nonumber\\
		& +\frac{1}{3\sqrt{5}} \bar{Y}_{20} \left[ -\left(F_x \bar{r}_x + F_y \bar{r}_y \right)+2 F_z \bar{r}_z \right] +\frac{1}{3} \bar{Y}_{00} \left[ \left(F_x \bar{r}_x + F_y \bar{r}_y \right)+ F_z \bar{r}_z \right] \nonumber\\
		= & Y_{00} \Bigg\{ F_x \bar{r}_x \left[\sqrt{\frac{2\pi}{15}} \left( \bar{Y}_{22} + \bar{Y}_{2-2}\right) - \sqrt{\frac{4\pi}{9}} \left( -\bar{Y}_{00} + \sqrt{\frac{1}{5}} \bar{Y}_{20}\right)\right] \nonumber\\
		& \qquad + F_y \bar{r}_y \left[-\sqrt{\frac{2\pi}{15}} \left( \bar{Y}_{22} + \bar{Y}_{2-2}\right) - \sqrt{\frac{4\pi}{9}} \left( -\bar{Y}_{00} + \sqrt{\frac{1}{5}} \bar{Y}_{20}\right)\right] \nonumber\\
		& \qquad + F_z \bar{r}_z \left[ \sqrt{\frac{4\pi}{9}} \left(\bar{Y}_{00}+ \sqrt{\frac{4}{5}} \bar{Y}_{20}\right) \right] \nonumber\\
		& \qquad + \left(F_x \bar{r}_y + F_y \bar{r}_x \right) \left[-i \sqrt{\frac{2\pi}{15}} \left( \bar{Y}_{22} - \bar{Y}_{2-2} \right)\right] \nonumber\\
		& \qquad + \left(F_x \bar{r}_z + F_z \bar{r}_x \right) \left[ -\sqrt{\frac{2\pi}{15}} \left( \bar{Y}_{21} - \bar{Y}_{2-1} \right)\right] \nonumber\\
		& \qquad + \left(F_y \bar{r}_z + F_y \bar{r}_x \right) \left[ i\sqrt{\frac{2\pi}{15}} \left( \bar{Y}_{21} - \bar{Y}_{2-1} \right)\right] \Bigg\} \nonumber\\
		= & Y_{00} \Big\{ F_x \bar{r}_x \bfhatb{r}_x \bfhatb{r}_x + F_y \bar{r}_y \bfhatb{r}_y \bfhatb{r}_y + F_z  \bar{r}_z  \bfhatb{r}_z  \bfhatb{r}_z \nonumber\\
		& \qquad + \left(F_x \bar{r}_y + F_y \bar{r}_x \right) \bfhatb{r}_x \bfhatb{r}_y + \left(F_x \bar{r}_z + F_z \bar{r}_x \right) \bfhatb{r}_x \bfhatb{r}_z + \left(F_y \bar{r}_z + F_z \bar{r}_y \right) \bfhatb{r}_y \bfhatb{r}_z\Big\} \nonumber\\
		= & Y_{00} \Big\{ F_x \bar{r}_x \underbrace{\left(\bfhatb{r}_x \bfhatb{r}_x + \bfhatb{r}_y \bfhatb{r}_y + \bfhatb{r}_z \bfhatb{r}_z\right)}_{1} + F_y \bar{r}_y \left(\bfhatb{r}_x \bfhatb{r}_x + \bfhatb{r}_y \bfhatb{r}_y + \bfhatb{r}_z \bfhatb{r}_z\right) \nonumber \\
		& \qquad + F_z \bar{r}_z \left(\bfhatb{r}_x \bfhatb{r}_x + \bfhatb{r}_y \bfhatb{r}_y + \bfhatb{r}_z \bfhatb{r}_z\right) \Big\} \nonumber\\
		= & Y_{00} \; \mbf{F} \cdot \mbfb{r}. \label{eq:G4_simpl0}
	\end{align}
	}
	Therefore, the term IV is zero for $n \geq 2$ because the first term and this term cancel.
	\newpage
	\item \begin{enumerate}[label=\alph*.]
		\item Same as for IV, but with positive global sign.
		\item Using Eq.~\eqref{eq:G4_simpl0} from IV again, we simplify the term for $n \geq 2$:
		\begin{align}
			& 2(n+2)\frac{n+3}{3} \mbfb{r} \cdot \mbf{F}\; Y_{00} + \frac{n}{2} (n+5) \left(Y_{00} \; \mbf{F} \cdot \mbfb{r} - \frac{1}{3} Y_{00} \; \mbf{F} \cdot \mbfb{r} \right) \nonumber\\
			=& Y_{00} \; \mbf{F} \cdot \mbfb{r} \left(\frac{2}{3} (n+2)(n+3) + \frac{1}{3} n(n+5)\right) \nonumber\\
			=& Y_{00} \; \mbf{F} \cdot \mbfb{r} \, (n+1)(n+4).
		\end{align}
	\end{enumerate}
	We can conclude from the above calculations that all terms that contribute to $u^{\bot}_{00}$ are proportional to $\mbf{F} \cdot \mbfb{r}$. This makes sense because the physical system as well as the spherical harmonic $Y_{00}$ that was used for the expansion are both isotropic. Thus, there is no preferred direction. Indeed, the deformational mode $u^{\bot}_{00}$ indicates isotropic volume expansion or contraction for positive and negative sign, respectively.
\end{enumerate}
	When summarizing all these terms into one equation, we get (we first list the contributions for $n=0$, then for $n=1$ and so on):
	{\allowdisplaybreaks
	\begin{alignat}{2}
		16\pi\mu(1-\nu) R^3 u^{\bot}_{00} &= 4\pi \; Y_{00} \; \mbf{F} \cdot \bfhatb{r} \Bigg\{&& \underbrace{\frac{2}{3} \beta_0 R \bar{r}}_{IV} - \underbrace{\frac{4}{3} \gamma_0 R \bar{r}}_{Va} + \underbrace{\frac{1}{3} k_1 R^2 \left( \frac{\bar{r}}{R} \right) }_{Ia} + \underbrace{4 \alpha_1 R \bar{r}}_{Vb} \nonumber\\
		& && + \!\!\!\sum_{\substack{n=1\\n \text{ odd}}}^{\infty} \Big[ \underbrace{\alpha_n R^2}_{Ib} + \underbrace{(2n+3) h_n R^2}_{IIa} + \underbrace{m_{n+2} \bar{r}^2}_{IIIa} \nonumber\\
		& && \qquad- \underbrace{\alpha_n R^2 \frac{1}{2}(n+1)(n+4)}_{IIb} 
		- \underbrace{\alpha_{n+2} \bar{r}^2 \frac{1}{2}(n+1)(n+4)}_{IIIb}\Big] \left( \frac{\bar{r}}{R} \right)^n \nonumber\\
		& && + \!\!\!\sum_{\substack{n=2\\n \text{ even}}}^{\infty} \!\! \Big[ -\underbrace{2 \gamma_n R \bar{r}}_{Va} + \underbrace{\alpha_{n+1} R \bar{r} (n+1)(n+4)}_{Ib} \Big] \left( \frac{\bar{r}}{R} \right)^n
		\Bigg\} \nonumber\\
		&= 4\pi \; Y_{00} \; \mbf{F} \cdot \bfhatb{r} \Bigg\{ && R \bar{r} \left[\frac{2}{3} \beta_0 - \frac{4}{3} \gamma_0 + \frac{1}{3} k_1  + 4 \alpha_1 -4 \alpha_1  + 5 h_1\right] \nonumber\\
		& && + \!\!\!\sum_{\substack{n=3\\n \text{ odd}}}^{\infty} \Big[ \alpha_n R^2 \left(1-\frac{1}{2}(n+1)(n+4)\right) 
		+ (2n+3) h_n R^2 + m_{n} R^2 \nonumber\\
		& && \qquad- \alpha_{n} R^2 \frac{1}{2}(n-1)(n+2)
		-2 \gamma_{n-1} R^2 + \alpha_{n} R^2 n(n+3) \Big] \left( \frac{\bar{r}}{R} \right)^n
		\Bigg\} \nonumber\\
		&= 4\pi \; Y_{00} \; \mbf{F} \cdot \bfhatb{r} \Bigg\{ && R \bar{r} \left[\frac{2}{3} \beta_0 - \frac{4}{3} \gamma_0 + \frac{1}{3} k_1  + 5 h_1\right] \Bigg\}. \label{eq:u00_1}
	\end{alignat}
	}
	In the last equation, we used the following two simplifications:
	\begin{align}
		\text{By definition: }2 \gamma_{n-1} = (2n+3)h_n+m_n, \label{eq:gamma_simpl}\\
		1 - \frac{1}{2}(n+1)(n+4) -\frac{1}{2}(n-1)(n+2) + n(n+3) = 0.
	\end{align}
	Inserting the following expressions according to our definitions
	\begin{align}
		\beta_0 &= \frac{1}{2} \left[ l_{1}-5h_{1} \right], \label{eq:beta_0}\\
		\gamma_0 &= \frac{1}{2} \left[ 5h_{1}+m_{1} \right], \label{eq:gamma_0}
	\end{align}
	leads to
	\begin{align}
		\frac{2}{3} \beta_0 - \frac{4}{3} \gamma_0 + \frac{1}{3} k_1  + 5 h_1 &= \frac{1}{3} \left( l_1 -2m_1 + k_1 \right). \label{eq:u00_2}
	\end{align}
	Finally, we combine Eq.~\eqref{eq:u00_1} with Eq.~\eqref{eq:u00_2}, insert the definitions from Eqs.~\eqref{eq:constants_1}--\eqref{eq:constants_14} and $Y_{00} = \sqrt{\frac{1}{4\pi}}$ and simplify (with the help of Mathematica \cite{Mathematica}) to
	\begin{align}
		u^{\bot}_{00} &= \mbf{F} \cdot \mbfb{r} \frac{\sqrt{4 \pi}}{16\pi\mu(1-\nu)\, R^2} \, \frac{2 \mu (1-\nu)(1-2\nu)}{2(1-2\nu) \tilde{\mu} + (1+\nu) \mu} \nonumber\\
		&= \mbf{F} \cdot \mbfb{r} \frac{1}{2 R^2} \, \frac{1-2\nu}{2(1-2\nu) \tilde{\mu} + (1+\nu) \mu} \sqrt{\frac{1}{4 \pi}}. \label{eq:u00_full}
	\end{align}
	We notice from Eq.~\eqref{eq:u00_full} that $u^{\bot}_{00}$ is always independent of the Poisson ratio outside the sphere $\tilde{\nu}$. Furthermore, in the limit of a free-standing sphere ($\tilde{\mu} \rightarrow 0$), we obtain
	\begin{align}
		u^{\bot}_{00} &= \mbf{F} \cdot \mbfb{r} \; \frac{1-2\nu}{1+\nu} \; \frac{1}{2 \mu R^2} \sqrt{\frac{1}{4 \pi}}. \label{eq:u00_manuscript}
	\end{align}
	We have confirmed that first taking the limit of $\tilde{\mu} \rightarrow 0$ (obtaining the solution for a free-standing sphere \cite{fischer2019magnetostriction}) and afterwards performing the same calculation as above leads to the same result as in Eq.~\eqref{eq:u00_manuscript}.

\newpage
\subsection{Calculation of $u^{\bot}_{20}$}
\label{subsec:u20}
Second, we evaluate the following mode of deformation:
\begin{align}
	u^{\bot}_{20} = \int_{0}^{\pi} d\theta\, \sin\theta \int_{0}^{2\pi}d\varphi \; \mbf{u}(\mbf{r}) \cdot \bfhat{r} \; Y_{20}.
\end{align}
In general, we proceed along similar lines as in the case for $u^{\bot}_{00}$, but the expressions here are more lengthy.
\begin{enumerate}[label=\Roman*.]
\item $ \tilde{G}_1 r^2 \mbf{F} $. \\
Projecting onto $ \bfhat{r} $ on the spherical surface of radius $ R $ leads to [see also Eq.~\eqref{eq:G1_sph2}]:
\begin{align}
	\tilde{G}_1 R^2 \mbf{F} \cdot \bfhat{r} &= \tilde{G}_1 R^2 \Bigg\{ -\sqrt{\frac{2\pi}{3}} \left[ F_x \left( Y_{11} - Y_{1-1}\right) -i\, F_y \left( Y_{11} + Y_{1-1}\right) \right]
	+F_z \sqrt{\frac{4\pi}{3}} Y_{10} \Bigg\}, 
	\label{eq:G1}\\
	\mbf{F} \cdot \bfhat{r}\; Y_{20} &= \sqrt{\frac{1}{5}}\Bigg\{ F_x \left( \sqrt{\frac{1}{6}} \left( Y_{11} - Y_{1-1}\right) - \sqrt{\frac{3}{7}} \left( Y_{31} - Y_{3-1}\right) \right) \nonumber \\
	&\qquad \quad -i\, F_y \left( \sqrt{\frac{1}{6}} \left( Y_{11} + Y_{1-1}\right) - \sqrt{\frac{3}{7}} \left( Y_{31} + Y_{3-1}\right) \right)
	+F_z \left[\sqrt{\frac{4}{3}}Y_{10} + \sqrt{\frac{9}{7}}Y_{30}\right]  \Bigg\}. \label{eq:G1_sph}
\end{align}
In the last equation, we used the identities from Eqs.~\eqref{eq:identities_1}--\eqref{eq:identities_3}.
\begin{enumerate}[label=\alph*.]
	\item $ k_n H^n $.
	Using the orthonormalization of the spherical harmonics and Eq.~\eqref{eq:add_sph}:
	\begin{align}
		&\int_{0}^{\pi} d\theta\, \sin\theta \int_{0}^{2\pi}d\varphi \; k_n H^n R^2 \mbf{F} \cdot \bfhat{r}\; Y_{20} \nonumber \\
		&\quad = \sqrt{\frac{1}{5}} k_n R^2\, 4\pi  \begin{cases}
			\frac{1}{3}\Bigg\{ \sqrt{\frac{1}{6}} \Bigg[ F_x \left( \bar{Y}_{11} - \bar{Y}_{1-1} \right) -i\, F_y \left( \bar{Y}_{11} + \bar{Y}_{1-1} \right) \Bigg]   \\
			\qquad +\sqrt{\frac{4}{3}}\, F_z\, \bar{Y}_{10} \Bigg\}, \quad n=1, \\
			\frac{1}{7}\Bigg\{ -\!\sqrt{\frac{3}{7}}\Bigg[ \, F_x  \left( \bar{Y}_{31} - \bar{Y}_{3-1} \right) -i\, F_y \left( \bar{Y}_{31} + \bar{Y}_{3-1}\right) \Bigg]  \\
			\qquad +\sqrt{\frac{9}{7}}\, F_z\, \bar{Y}_{30}  \Bigg\}, \quad n=3, \\
			0, \quad \text{else.}
		\end{cases}
		\label{eq:G1a}
	\end{align}
	\item $ (h_n-e_n-u_{n+2}) K^n = \alpha_n K^n$.
	Again, using the orthonormalization of the spherical harmonics, Eq.~\eqref{eq:add_sph}, and this time also Eq.~\eqref{eq:deriv1}:
	\begin{align}
		&\int_{0}^{\pi} d\theta\, \sin\theta \int_{0}^{2\pi}d\varphi \; \alpha_n K^n R^2 \mbf{F} \cdot \bfhat{r}\; Y_{20} \nonumber \\
		&\quad = \sqrt{\frac{1}{5}} \alpha_n R^2\, 4\pi \begin{cases}
			\sqrt{\frac{1}{6}} \Bigg[ F_x \left( \bar{Y}_{11} - \bar{Y}_{1-1}\right) -i\, F_y \left( \bar{Y}_{11} + \bar{Y}_{1-1} \right) \Bigg] 
			+\sqrt{\frac{4}{3}}\, F_z\, \bar{Y}_{10}, \quad n=1,\\
			\sqrt{\frac{1}{6}} \Bigg[ F_x \left( \bar{Y}_{11} - \bar{Y}_{1-1}\right) -i\, F_y \left( \bar{Y}_{11} + \bar{Y}_{1-1}\right) \Bigg]  \\
			\qquad -\sqrt{\frac{3}{7}}\Bigg[ \, F_x  \left( \bar{Y}_{31} - \bar{Y}_{3-1}\right) -i\, F_y \left( \bar{Y}_{31} + \bar{Y}_{3-1}\right) \Bigg]   \\
			\qquad +F_z \left[\sqrt{\frac{4}{3}}\bar{Y}_{10} + \sqrt{\frac{9}{7}}\bar{Y}_{30}\right],
			\quad n \geq 3 \text{ odd}, \\
			0, \quad \text{else.}
		\end{cases} \label{eq:G1b}
	\end{align}
\end{enumerate}
\newpage
\item $ \tilde{G}_2 (\mbf{r}\mbf{r}) \cdot \mbf{F} $. \\
\begin{enumerate}[label=\alph*.]
	\item $ (2n+3)h_n K^n $.
	The result is the same as in Eq.~\eqref{eq:G1b}, replacing $ \alpha_n $ by $ (2n+3)h_n $.
	\item $ (-h_n+e_n+u_{n+2}) L^n = -\alpha_n L^n $.
	\begin{align}
		&-\int_{0}^{\pi} d\theta\, \sin\theta \int_{0}^{2\pi}d\varphi \; \alpha_n L^n R^2 \mbf{F} \cdot \bfhat{r}\; Y_{20} \nonumber \\
		&\quad = -\sqrt{\frac{1}{5}} \alpha_n R^2\, 4\pi  \begin{cases}
			5\Bigg\{ \sqrt{\frac{1}{6}} \Bigg[ F_x \left( \bar{Y}_{11} - \bar{Y}_{1-1}\right) -i\, F_y \left( \bar{Y}_{11} + \bar{Y}_{1-1}\right) \Bigg] +\sqrt{\frac{4}{3}}\, F_z\, \bar{Y}_{10} \Bigg\}, \quad n=1, \\
			\Bigg\{ \sqrt{\frac{1}{6}} \Bigg[ F_x \left( \bar{Y}_{11} - \bar{Y}_{1-1}\right) -i\, F_y \left( \bar{Y}_{11} + \bar{Y}_{1-1} \right) \Bigg]
			+\sqrt{\frac{4}{3}}\, F_z\, \bar{Y}_{10} \Bigg\} \frac{n+1}{2}(n+4)\\
			+\Bigg\{ -\!\sqrt{\frac{3}{7}}\Bigg[ \, F_x  \left( \bar{Y}_{31} - \bar{Y}_{3-1}\right) -i\, F_y \left( \bar{Y}_{31} + \bar{Y}_{3-1}\right) \Bigg]  \\
			\qquad +\sqrt{\frac{9}{7}}\, F_z\, \bar{Y}_{30}  \Bigg\} \frac{n-1}{2}(n+6),\quad n \geq 3 \text{ odd}, \\
			0, \quad \text{else.}
		\end{cases}
		\label{eq:G2b}
	\end{align}
\end{enumerate}
\item $ \tilde{G}_3 (\mbfb{r}\mbfb{r}) \cdot \mbf{F} $. \\
Similar to Eq.~\eqref{eq:G1_sph}:
\begin{align}
	\mbfb{r} \cdot \bfhat{r}\; Y_{20} &= \sqrt{\frac{1}{5}}\Bigg\{ \Bigg[ \bar{r}_x \left( \sqrt{\frac{1}{6}} \left( Y_{11} - Y_{1-1}\right) - \sqrt{\frac{3}{7}} \left( Y_{31} - Y_{3-1}\right) \right) \nonumber \\
	&\qquad \quad -i\, \bar{r}_y \left( \sqrt{\frac{1}{6}} \left( Y_{11} + Y_{1-1}\right) - \sqrt{\frac{3}{7}} \left( Y_{31} + Y_{3-1}\right) \right) \Bigg] 
	+\bar{r}_z \left[\sqrt{\frac{4}{3}}Y_{10} + \sqrt{\frac{9}{7}}Y_{30}\right]  \Bigg\}. \label{eq:G3_sph}
\end{align}
\begin{enumerate}[label=\alph*.]
	\item $ m_{n+2} K^n $.
	The result is similar to Eq.~\eqref{eq:G1b}:
	\begin{align}
		&\int_{0}^{\pi} d\theta\, \sin\theta \int_{0}^{2\pi}d\varphi \; m_{n+2} K^n \; \mbfb{r} \cdot \mbf{F} \; \mbfb{r} \cdot \bfhat{r}\; Y_{20} \nonumber \\
		&\quad = \sqrt{\frac{1}{5}} m_{n+2} \mbfb{r} \cdot \mbf{F}\; 4\pi \begin{cases}
			\sqrt{\frac{1}{6}} \Bigg[ \bar{r}_x \left( \bar{Y}_{11} - \bar{Y}_{1-1}\right) -i\, \bar{r}_y \left( \bar{Y}_{11} + \bar{Y}_{1-1} \right) \Bigg]
			+\sqrt{\frac{4}{3}}\, \bar{r}_z\, \bar{Y}_{10}, \quad n=1,\\
			\sqrt{\frac{1}{6}} \Bigg[ \bar{r}_x \left( \bar{Y}_{11} - \bar{Y}_{1-1}\right) -i\, \bar{r}_y \left( \bar{Y}_{11} + \bar{Y}_{1-1}\right) \Bigg]  \\
			\qquad -\sqrt{\frac{3}{7}}\Bigg[ \, \bar{r}_x  \left( \bar{Y}_{31} - \bar{Y}_{3-1}\right) -i\, \bar{r}_y \left( \bar{Y}_{31} + \bar{Y}_{3-1}\right) \Bigg]   \\
			\qquad +\bar{r}_z \left[\sqrt{\frac{4}{3}}\bar{Y}_{10} + \sqrt{\frac{9}{7}}\bar{Y}_{30}\right],
			\quad n \geq 3 \text{ odd}, \\
			0, \quad \text{else.}
		\end{cases} \label{eq:G3a}
	\end{align}
	\item $ (-h_{n+2}+e_{n+2}+u_{n+4}) L^n = -\alpha_{n+2} L^n $.
	The result is similar to Eq.~\eqref{eq:G2b}:
	\begin{align}
		&-\int_{0}^{\pi} d\theta\, \sin\theta \int_{0}^{2\pi}d\varphi \; \alpha_{n+2} L^n \; \mbfb{r} \cdot \mbf{F} \; \mbfb{r} \cdot \bfhat{r}\; Y_{20} \nonumber \\
		&\quad = -\sqrt{\frac{1}{5}} \alpha_{n+2} \mbfb{r} \cdot \mbf{F}\; 4\pi  \begin{cases}
			5\Bigg\{ \sqrt{\frac{1}{6}} \Bigg[ \bar{r}_x \left( \bar{Y}_{11} - \bar{Y}_{1-1}\right) -i\, \bar{r}_y \left( \bar{Y}_{11} + \bar{Y}_{1-1}\right) \Bigg]
			+\sqrt{\frac{4}{3}}\, \bar{r}_z\, \bar{Y}_{10} \Bigg\}, \quad n=1, \\
			\Bigg\{ \sqrt{\frac{1}{6}} \Bigg[ \bar{r}_x \left( \bar{Y}_{11} - \bar{Y}_{1-1}\right) -i\, \bar{r}_y \left( \bar{Y}_{11} + \bar{Y}_{1-1} \right) \Bigg] 
			+\sqrt{\frac{4}{3}}\, \bar{r}_z\, \bar{Y}_{10} \Bigg\} \frac{n+1}{2}(n+4)\\
			+\Bigg\{ -\!\sqrt{\frac{3}{7}}\Bigg[ \, \bar{r}_x  \left( \bar{Y}_{31} - \bar{Y}_{3-1}\right) -i\, \bar{r}_y \left( \bar{Y}_{31} + \bar{Y}_{3-1}\right) \Bigg]  \\
			\qquad +\sqrt{\frac{9}{7}}\, \bar{r}_z\, \bar{Y}_{30}  \Bigg\} \frac{n-1}{2}(n+6),\quad n \geq 3 \text{ odd}, \\
			0, \quad \text{else.}
		\end{cases}
		\label{eq:G3b}
	\end{align}
\end{enumerate}
\item $ \tilde{G}_4 (\mbf{r}\mbfb{r}-\mbfb{r}\mbf{r}) \cdot \mbf{F} $. \\
Projecting onto $ \bfhat{r} $ on the spherical surface of radius $ R $ leads to  [see Eq.~\eqref{eq:G4_1} again and the equations thereafter]:
\begin{align}
	\mbfb{r} \cdot \mbf{F} \; Y_{20} &= \mbfb{r} \cdot \mbf{F} \; Y_{20} \text{ (already in final form)}, \label{eq:G4_2}\\
	\mbf{F} \cdot \bfhat{r}\; \bfhat{r} \cdot \mbfb{r}\; Y_{20} &= F_x \bar{r}_x \Bigg[ \sqrt{\frac{2\pi}{15}} \left( Y_{22} + Y_{2-2} \right) - \sqrt{\frac{4\pi}{9}} \left( -Y_{00} + \sqrt{\frac{1}{5}} Y_{20} \right)\Bigg] Y_{20} \nonumber \\
	&\quad + \left(F_x \bar{r}_y + F_y \bar{r}_x \right) \Bigg[-i \sqrt{\frac{2\pi}{15}} \left( Y_{22} - Y_{2-2} \right)\Bigg] Y_{20} \nonumber \\
	&\quad +F_y \bar{r}_y \Bigg[-\sqrt{\frac{2\pi}{15}} \left( Y_{22} + Y_{2-2} \right) - \sqrt{\frac{4\pi}{9}} \left( -Y_{00} + \sqrt{\frac{1}{5}} Y_{20} \right)\Bigg] Y_{20} \nonumber \\
	&\quad + \left(F_x \bar{r}_z + F_z \bar{r}_x \right) \Bigg[- \sqrt{\frac{2\pi}{15}} \left(Y_{21} - Y_{2-1} \right)\Bigg] Y_{20} \nonumber \\
	&\quad + \left(F_y \bar{r}_z + F_z \bar{r}_y \right) \Bigg[i \sqrt{\frac{2\pi}{15}} \left(Y_{21} + Y_{2-1} \right)\Bigg] Y_{20} \nonumber \\
	&\quad +F_z \bar{r}_z \Bigg[\sqrt{\frac{4\pi}{9}} \left(Y_{00} + \sqrt{\frac{4}{5}} Y_{20} \right)\Bigg] Y_{20} \nonumber\\
	&= \sqrt{\frac{2\pi}{15}} \Big[ \left( Y_{22} + Y_{2-2} \right) \left(F_x \bar{r}_x - F_y \bar{r}_y \right) 
	-i \left( Y_{22} - Y_{2-2} \right) \left(F_x \bar{r}_y + F_y \bar{r}_x \right) \nonumber \\
	&\qquad \qquad -\left(Y_{21} - Y_{2-1} \right) \left(F_x \bar{r}_z + F_z \bar{r}_x \right) \nonumber +i \left(Y_{21} + Y_{2-1} \right) \left(F_y \bar{r}_z + F_z \bar{r}_y \right) \Big] Y_{20}, \nonumber \\
	&\quad + \sqrt{\frac{4\pi}{9}} \Bigg[ \left( Y_{00} - \sqrt{\frac{1}{5}} Y_{20} \right) \left(F_x \bar{r}_x + F_y \bar{r}_y \right) 
	+ \left(Y_{00} + \sqrt{\frac{4}{5}} Y_{20} \right) F_z \bar{r}_z \Bigg] Y_{20} \end{align}
	Using the identities from Eqs.~\eqref{eq:identities_4}--\eqref{eq:identities_8}, we rewrite:
	\begin{align}
	\mbf{F} \cdot \bfhat{r}\; \bfhat{r} \cdot \mbfb{r}\; Y_{20} &= \frac{1}{7\sqrt{6}} \Big[ \left(-2\left( Y_{22} + Y_{2-2} \right) + \sqrt{3} \left( Y_{42} + Y_{4-2} \right) \right) \left(F_x \bar{r}_x - F_y \bar{r}_y \right) \nonumber \\
	&\quad -i \left(-2\left( Y_{22} - Y_{2-2} \right) + \sqrt{3} \left( Y_{42} - Y_{4-2} \right) \right) \left(F_x \bar{r}_y + F_y \bar{r}_x \right) \nonumber \\
	&\quad -\left(Y_{21} - Y_{2-1} +\sqrt{6} \left(Y_{41} - Y_{4-1}\right) \right) \left(F_x \bar{r}_z + F_z \bar{r}_x \right) \nonumber \\
	&\quad +i \left(Y_{21} + Y_{2-1} +\sqrt{6} \left(Y_{41} + Y_{4-1}\right) \right) \left(F_y \bar{r}_z + F_z \bar{r}_y \right) \Big]\nonumber \\
	&\quad + \frac{1}{3\sqrt{5}} \Bigg[ -\left( Y_{00} - \frac{5\sqrt{5}}{7} Y_{20} + \frac{6}{7} Y_{40} \right) \left(F_x \bar{r}_x + F_y \bar{r}_y \right) \nonumber \\
	&\quad +\left( 2\, Y_{00} + \frac{11\sqrt{5}}{7} Y_{20} + \frac{12}{7} Y_{40}\right) F_z \bar{r}_z \Bigg]. \label{eq:G4_3}
\end{align}
Only $ \beta_n K^n $ contributes to $\tilde{G}_4$:
\begin{align}
	&\int_{0}^{\pi} d\theta\, \sin\theta \int_{0}^{2\pi}d\varphi \; \beta_n K^n R\; \left(\mbfb{r} \cdot \mbf{F} - \mbf{F} \cdot \bfhat{r}\; \bfhat{r} \cdot \mbfb{r}\right) \, Y_{20} \nonumber \\
	&\quad = \beta_n R \; 4\pi \begin{cases}
		\frac{-1}{3\sqrt{5}} \bar{Y}_{00} \left[ -\left(F_x \bar{r}_x + F_y \bar{r}_y \right)+2 F_z \bar{r}_z \right], \quad n=0,\\
		\frac{-1}{3\sqrt{5}} \bar{Y}_{00} \left[ -\left(F_x \bar{r}_x + F_y \bar{r}_y \right)+2 F_z \bar{r}_z \right]\\
		\qquad +\mbfb{r} \cdot \mbf{F}\; \bar{Y}_{20} 
		- \frac{1}{7\sqrt{6}} \Big[ -2\left( \bar{Y}_{22} + \bar{Y}_{2-2} \right) \left(F_x \bar{r}_x - F_y \bar{r}_y \right)\\
		\qquad +2i \left( \bar{Y}_{22} - \bar{Y}_{2-2} \right)\left(F_x \bar{r}_y + F_y \bar{r}_x \right) 
		-\left(\bar{Y}_{21} - \bar{Y}_{2-1} \right) \left(F_x \bar{r}_z + F_z \bar{r}_x \right)\\
		\qquad + i \left(\bar{Y}_{21} + \bar{Y}_{2-1} \right) \left(F_y \bar{r}_z + F_z \bar{r}_y \right) \Big]\\
		\qquad -\frac{1}{21} \bar{Y}_{20} \Big[ 5\left(F_x \bar{r}_x + F_y \bar{r}_y \right)+11\, F_z \bar{r}_z \Big],	\quad n = 2,\\
		\frac{-1}{3\sqrt{5}} \bar{Y}_{00} \left[ -\left(F_x \bar{r}_x + F_y \bar{r}_y \right)+2 F_z \bar{r}_z \right]\\
		\qquad +\mbfb{r} \cdot \mbf{F}\; \bar{Y}_{20} - \frac{1}{7\sqrt{6}} \Big[ -2\left( \bar{Y}_{22} + \bar{Y}_{2-2} \right) \left(F_x \bar{r}_x - F_y \bar{r}_y \right)\\
		\qquad +2i \left( \bar{Y}_{22} - \bar{Y}_{2-2} \right)\left(F_x \bar{r}_y + F_y \bar{r}_x \right) 
		-\left(\bar{Y}_{21} - \bar{Y}_{2-1} \right) \left(F_x \bar{r}_z + F_z \bar{r}_x \right)\\
		\qquad + i \left(\bar{Y}_{21} + \bar{Y}_{2-1} \right) \left(F_y \bar{r}_z + F_z \bar{r}_y \right) \Big]
		-\frac{1}{21} \bar{Y}_{20} \Big[ 5\left(F_x \bar{r}_x + F_y \bar{r}_y \right)+11\, F_z \bar{r}_z \Big]\\
		\qquad - \frac{1}{7\sqrt{2}} \Big[ \left( \bar{Y}_{42} + \bar{Y}_{4-2} \right) \left(F_x \bar{r}_x - F_y \bar{r}_y \right)
		-i \left( \bar{Y}_{42} - \bar{Y}_{4-2} \right) \left(F_x \bar{r}_y + F_y \bar{r}_x \right)\\
		\qquad -\sqrt{2} \left(\bar{Y}_{41} - \bar{Y}_{4-1}\right) \left(F_x \bar{r}_z + F_z \bar{r}_x \right) 
		+\sqrt{2}i \left(\bar{Y}_{41} + \bar{Y}_{4-1}\right) \left(F_y \bar{r}_z + F_z \bar{r}_y \right) \Big]\\
		\qquad - \frac{2}{7\sqrt{5}} \bar{Y}_{40} \Big[ -\left(F_x \bar{r}_x + F_y \bar{r}_y \right)+2\, F_z \bar{r}_z \Big], \quad n \geq 4 \text{ even}, \\
		0, \quad \text{else.}
	\end{cases} \label{eq:G4}
\end{align}
This can be simplified slightly using the following identity:
\begin{align}
	\mbfb{r} \cdot \mbf{F} -\frac{1}{21} \Big[ 5\left(F_x \bar{r}_x + F_y \bar{r}_y \right)+11\, F_z \bar{r}_z \Big] &= \frac{2}{21} \Big[ 8\left(F_x \bar{r}_x + F_y \bar{r}_y \right)+5\, F_z \bar{r}_z \Big].
	\label{eq:G4_simpl}
\end{align}
\newpage
\item $ \tilde{G}_5 (\mbf{r}\mbfb{r}+\mbfb{r}\mbf{r}) \cdot \mbf{F} $. \\
Here, we can use Eq.~\eqref{eq:G4_1}, but with a plus instead of a minus. Thus, also Eq.~\eqref{eq:G4_2} and Eq.~\eqref{eq:G4_3} can be used again.
\begin{enumerate}[label=\alph*.]
	\item $ - \gamma_n K^n $.\\
	Let $ \gamma_n \coloneqq \frac{1}{2} \left[ (2n+5)h_{n+1} + m_{n+1} \right] $.
	\begin{align}
		&-\int_{0}^{\pi} d\theta\, \sin\theta \int_{0}^{2\pi}d\varphi \; \gamma_n K^n R\; \left(\mbfb{r} \cdot \mbf{F} + \mbf{F} \cdot \bfhat{r}\; \bfhat{r} \cdot \mbfb{r}\right) \, Y_{20} \nonumber \\
		&\quad = -\gamma_n R \; 4\pi \begin{cases}
			\frac{1}{3\sqrt{5}} \bar{Y}_{00} \left[ -\left(F_x \bar{r}_x + F_y \bar{r}_y \right)+2 F_z \bar{r}_z \right], \quad n=0,\\
			\frac{1}{3\sqrt{5}} \bar{Y}_{00} \left[ -\left(F_x \bar{r}_x + F_y \bar{r}_y \right)+2 F_z \bar{r}_z \right]\\
			\qquad +\mbfb{r} \cdot \mbf{F}\; \bar{Y}_{20} + \frac{1}{7\sqrt{6}} \Big[ -2\left( \bar{Y}_{22} + \bar{Y}_{2-2} \right) \left(F_x \bar{r}_x - F_y \bar{r}_y \right)\\
			\qquad +2i \left( \bar{Y}_{22} - \bar{Y}_{2-2} \right)\left(F_x \bar{r}_y + F_y \bar{r}_x \right)
			-\left(\bar{Y}_{21} - \bar{Y}_{2-1} \right) \left(F_x \bar{r}_z + F_z \bar{r}_x \right)\\
			\qquad + i \left(\bar{Y}_{21} + \bar{Y}_{2-1} \right) \left(F_y \bar{r}_z + F_z \bar{r}_y \right) \Big]\\
			\qquad +\frac{1}{21} \bar{Y}_{20} \Big[ 5\left(F_x \bar{r}_x + F_y \bar{r}_y \right)+11\, F_z \bar{r}_z \Big],	\quad n = 2,\\
			\frac{1}{3\sqrt{5}} \bar{Y}_{00} \left[ -\left(F_x \bar{r}_x + F_y \bar{r}_y \right)+2 F_z \bar{r}_z \right]\\
			\qquad +\mbfb{r} \cdot \mbf{F}\; \bar{Y}_{20} + \frac{1}{7\sqrt{6}} \Big[ -2\left( \bar{Y}_{22} + \bar{Y}_{2-2} \right) \left(F_x \bar{r}_x - F_y \bar{r}_y \right)\\
			\qquad +2i \left( \bar{Y}_{22} - \bar{Y}_{2-2} \right)\left(F_x \bar{r}_x + F_y \bar{r}_y \right)
			-\left(\bar{Y}_{21} - \bar{Y}_{2-1} \right) \left(F_x \bar{r}_z + F_z \bar{r}_x \right)\\
			\qquad + i \left(\bar{Y}_{21} + \bar{Y}_{2-1} \right) \left(F_y \bar{r}_z + F_z \bar{r}_y \right) \Big]
			+\frac{1}{21} \bar{Y}_{20} \Big[ 5\left(F_x \bar{r}_x + F_y \bar{r}_y \right)+11\, F_z \bar{r}_z \Big]\\
			\qquad + \frac{1}{7\sqrt{2}} \Big[ \left( \bar{Y}_{42} + \bar{Y}_{4-2} \right) \left(F_x \bar{r}_x - F_y \bar{r}_y \right)
			-i \left( \bar{Y}_{42} - \bar{Y}_{4-2} \right) \left(F_x \bar{r}_y + F_y \bar{r}_x \right)\\
			\qquad -\sqrt{2} \left(\bar{Y}_{41} - \bar{Y}_{4-1}\right) \left(F_x \bar{r}_z + F_z \bar{r}_x \right) +\sqrt{2}i \left(\bar{Y}_{41} + \bar{Y}_{4-1}\right) \left(F_y \bar{r}_z + F_z \bar{r}_y \right) \Big]\\
			\qquad + \frac{2}{7\sqrt{5}} \bar{Y}_{40} \Big[ -\left(F_x \bar{r}_x + F_y \bar{r}_y \right)+2\, F_z \bar{r}_z \Big], \quad n \geq 4 \text{ even}, \\
			0, \quad \text{else.}
		\end{cases} \label{eq:G5a}
	\end{align}
	Similar to Eq.~\eqref{eq:G4_simpl}, we simplify as follows:
	\begin{align}
		\mbfb{r} \cdot \mbf{F} +\frac{1}{21} \Big[ 5\left(F_x \bar{r}_x + F_y \bar{r}_y \right)+11\, F_z \bar{r}_z \Big] &= \frac{2}{21} \Big[ 13 \left(F_x \bar{r}_x + F_y \bar{r}_y \right) + 16 F_z \bar{r}_z \Big].
		\label{eq:G5_simpl}
	\end{align}
	\item $ (h_{n+1}-e_{n+1}-u_{n+3}) L^n = \alpha_{n+1} L^n $.
	\begin{align}
		&\int_{0}^{\pi} d\theta\, \sin\theta \int_{0}^{2\pi}d\varphi \; \alpha_{n+1} L^n R\; \left(\mbfb{r} \cdot \mbf{F} + \mbf{F} \cdot \bfhat{r}\; \bfhat{r} \cdot \mbfb{r}\right) \, Y_{20} \nonumber \\
		&\quad = \alpha_{n+1} R \; 4\pi \begin{cases}
			\frac{1}{\sqrt{5}} \bar{Y}_{00} \left[ -\left(F_x \bar{r}_x + F_y \bar{r}_y \right)+2 F_z \bar{r}_z \right], \quad n=0,\\
			\frac{10}{3\sqrt{5}} \bar{Y}_{00} \left[ -\left(F_x \bar{r}_x + F_y \bar{r}_y \right)+2 F_z \bar{r}_z \right]\\
			\qquad + \frac{1}{\sqrt{6}} \Big[ -2\left( \bar{Y}_{22} + \bar{Y}_{2-2} \right) \left(F_x \bar{r}_x - F_y \bar{r}_y \right)\\
			\qquad +2i \left( \bar{Y}_{22} - \bar{Y}_{2-2} \right)\left(F_x \bar{r}_y + F_y \bar{r}_x \right) 
			-\left(\bar{Y}_{21} - \bar{Y}_{2-1} \right) \left(F_x \bar{r}_z + F_z \bar{r}_x \right)\\
			\qquad + i \left(\bar{Y}_{21} + \bar{Y}_{2-1} \right) \left(F_y \bar{r}_z + F_z \bar{r}_y \right) \Big]\\
			\qquad +\frac{2}{3} \bar{Y}_{20} \Big[ 13\left(F_x \bar{r}_x + F_y \bar{r}_y \right)+16\, F_z \bar{r}_z \Big],	\quad n = 2,\\
			\frac{n+2}{2}\frac{n+3}{3\sqrt{5}} \bar{Y}_{00} \left[ -\left(F_x \bar{r}_x + F_y \bar{r}_y \right)+2 F_z \bar{r}_z \right]\\
			\qquad + \frac{n}{2} \frac{n+5}{7\sqrt{6}} \Big[ -2\left( \bar{Y}_{22} + \bar{Y}_{2-2} \right) \left(F_x \bar{r}_x - F_y \bar{r}_y \right)\\
			\qquad +2i \left( \bar{Y}_{22} - \bar{Y}_{2-2} \right)\left(F_x \bar{r}_y + F_y \bar{r}_x \right) 
			-\left(\bar{Y}_{21} - \bar{Y}_{2-1} \right) \left(F_x \bar{r}_z + F_z \bar{r}_x \right)\\
			\qquad + i \left(\bar{Y}_{21} + \bar{Y}_{2-1} \right) \left(F_y \bar{r}_z + F_z \bar{r}_y \right) \Big]
			+n \frac{n+5}{21} \bar{Y}_{20} \Big[ 13\left(F_x \bar{r}_x + F_y \bar{r}_y \right)+16\, F_z \bar{r}_z \Big]\\
			\qquad + \frac{n-2}{2}\frac{n+7}{7\sqrt{2}} \Big[ \left( \bar{Y}_{42} + \bar{Y}_{4-2} \right) \left(F_x \bar{r}_x - F_y \bar{r}_y \right)
			-i \left( \bar{Y}_{42} - \bar{Y}_{4-2} \right) \left(F_x \bar{r}_x + F_y \bar{r}_y \right)\\
			\qquad -\sqrt{2} \left(\bar{Y}_{41} - \bar{Y}_{4-1}\right) \left(F_x \bar{r}_z + F_z \bar{r}_x \right)+\sqrt{2}i \left(\bar{Y}_{41} + \bar{Y}_{4-1}\right) \left(F_y \bar{r}_z + F_z \bar{r}_y \right) \Big]\\
			\qquad + \left(n-2\right) \frac{n+7}{7\sqrt{5}} \bar{Y}_{40} \Big[ -\left(F_x \bar{r}_x + F_y \bar{r}_y \right)+2\, F_z \bar{r}_z \Big], \quad n \geq 4 \text{ even}, \\
			0, \quad \text{else.}
		\end{cases} \label{eq:G5b}
	\end{align}
\end{enumerate}

\end{enumerate}
\noindent
Again, the terms I--V encompass all contributions to $ u^{\bot}_{20} $. Thus, we can get the final result for $ u^{\bot}_{20} $ by summing up all contributions from I--V and by adding terms for all $ n \in \mathbb{N}_0 $ (with a prefactor of $ \left(\frac{\bar{r}}{r}\right)^n $). Furthermore, the whole result needs to be divided by $ 16\pi\mu(1-\nu) R^3$.\\

Next, we simplify the terms I--V again by applying Eqs.~\eqref{eq:r_x}--\eqref{eq:r_z} in reverse, but for $\mbfb{r}$ (replacing $\bfhat{r}$ by $\bfhatb{r}$ and $Y_{lm}$ by $\bar{Y}_{lm}$):
\begin{enumerate}[label=\Roman*.]
	\item Terms for $n=1$ (for a and b):
	\begin{align}
		\sqrt{\frac{1}{6}} \Bigg[ F_x \left( \bar{Y}_{11} - \bar{Y}_{1-1} \right) -i\, F_y \left( \bar{Y}_{11} + \bar{Y}_{1-1} \right) \Bigg]
		+\sqrt{\frac{4}{3}}\, F_z\, \bar{Y}_{10}
		= Y_{00} \left[-F_x \bfhatb{r}_x - F_y \bfhatb{r}_y + 2 F_z \bfhatb{r}_z\right]. \label{eq:G1_ab3}
	\end{align}
	Terms for $n=3$ (for a and b):
	First, we need the following identities, similar to those from Eq.~\eqref{eq:r_x}--\eqref{eq:rr_zz}:
	\begin{align}
		\bar{Y}_{31} - \bar{Y}_{3-1} &= \frac{1}{4} \sqrt{\frac{21}{\pi}}\; \bfhatb{r}_x \left(1-5\bfhatb{r}_z\bfhatb{r}_z\right) \\
		-i\left(\bar{Y}_{31} + \bar{Y}_{3-1}\right) &= \frac{1}{4} \sqrt{\frac{21}{\pi}}\; \bfhatb{r}_y \left(1-5\bfhatb{r}_z\bfhatb{r}_z\right) \\
		\bar{Y}_{30} &= \frac{1}{4} \sqrt{\frac{7}{\pi}}\; \bfhatb{r}_z \left(-3+5\bfhatb{r}_z\bfhatb{r}_z\right).
	\end{align}
	Therefore, we can simplify the expression for $n=3$ as
	\begin{align}
		& -\sqrt{\frac{3}{7}}\Bigg[ \, F_x  \left( \bar{Y}_{31} - \bar{Y}_{3-1} \right) -i\, F_y \left( \bar{Y}_{31} + \bar{Y}_{3-1}\right) \Bigg] +\sqrt{\frac{9}{7}}\, F_z\, \bar{Y}_{30} \nonumber\\
		= & \frac{3}{4\sqrt{\pi}} \left[-F_x \bfhatb{r}_x - F_y \bfhatb{r}_y - 3 F_z \bfhatb{r}_z +5 \bfhatb{r}_z \bfhatb{r}_z \left(F_x \bfhatb{r}_x + F_y \bfhatb{r}_y + F_z \bfhatb{r}_z \right)\right] \nonumber\\
		= & \frac{3}{2} Y_{00} \left[-F_x \bfhatb{r}_x - F_y \bfhatb{r}_y - 3 F_z \bfhatb{r}_z +5 \bfhatb{r}_z \bfhatb{r}_z \, \mbf{F} \cdot \bfhatb{r} \right]. \label{eq:G1_ab4}
	\end{align}
	\item Here, we simplify the terms in the same manner as in I because only the prefactors are changed in II.
	\item Similar expression for $n=1$ as in Eq.~\eqref{eq:G1_ab3} (for a and b):
	\begin{align}
		\sqrt{\frac{1}{6}} \Bigg[ \bar{r}_x \left( \bar{Y}_{11} - \bar{Y}_{1-1} \right) -i\, \bar{r}_y \left( \bar{Y}_{11} + \bar{Y}_{1-1} \right) \Bigg]
		+\sqrt{\frac{4}{3}}\, \bar{r}_z\, \bar{Y}_{10} = & Y_{00} \left[-\bar{r}_x \bfhatb{r}_x - \bar{r}_y \bfhatb{r}_y + 2 \bar{r}_z \bfhatb{r}_z\right] \nonumber\\
		= & Y_{00} \, \bar{r} \left( -1 + 3 \bfhatb{r}_z \bfhatb{r}_z \right). \label{eq:G3_ab3}
	\end{align}
	Analogously, the terms for $n \geq 3$ are similar to Eq.~\eqref{eq:G1_ab4}:
	\begin{align}
		-\sqrt{\frac{3}{7}}\Bigg[ \, \bar{r}_x  \left( \bar{Y}_{31} - \bar{Y}_{3-1} \right) -i\, \bar{r}_y \left( \bar{Y}_{31} + \bar{Y}_{3-1}\right) \Bigg] +\sqrt{\frac{9}{7}}\, \bar{r}_z\, \bar{Y}_{30} 
		= & \frac{3}{2} Y_{00} \left[-\bar{r}_x \bfhatb{r}_x - \bar{r}_y \bfhatb{r}_y - 3 \bar{r}_z \bfhatb{r}_z +5 \bfhatb{r}_z \bfhatb{r}_z \, \mbfb{r} \cdot \bfhatb{r} \right] \nonumber\\
		= & \frac{3}{2} Y_{00} \, \bar{r} \left( -1 + 3 \bfhatb{r}_z \bfhatb{r}_z \right). \label{eq:G3_ab4}
	\end{align}
	\newpage
	\item The term for $n=0$ is already simplified as far as possible. The new part of the term for $n=2$ (compared to $n=0$) can be simplified using the equations from Eqs.~\eqref{eq:rr_xx}--\eqref{eq:rr_zz} (omitting the global negative sign):
	\begin{align}
		&\frac{1}{7\sqrt{6}} \Big[ -2\left( \bar{Y}_{22} + \bar{Y}_{2-2} \right) \left(F_x \bar{r}_x - F_y \bar{r}_y \right)+2i \left( \bar{Y}_{22} - \bar{Y}_{2-2} \right)\left(F_x \bar{r}_y + F_y \bar{r}_x \right) \nonumber\\
		&\qquad -\left(\bar{Y}_{21} - \bar{Y}_{2-1} \right) \left(F_x \bar{r}_z + F_z \bar{r}_x \right) + i \left(\bar{Y}_{21} + \bar{Y}_{2-1} \right) \left(F_y \bar{r}_z + F_z \bar{r}_y \right) \Big] \nonumber\\
		& +\frac{1}{21} \bar{Y}_{20} \left[ 5\left(F_x \bar{r}_x + F_y \bar{r}_y \right)+11\, F_z \bar{r}_z \right] - \mbfb{r} \cdot \mbf{F}\; \bar{Y}_{20}\nonumber\\
		= & \frac{\sqrt{5}}{7} Y_{00} \Bigg[ -\left( \bfhatb{r}_x \bfhatb{r}_x - \bfhatb{r}_y \bfhatb{r}_y \right) \left(F_x \bar{r}_x - F_y \bar{r}_y \right)  -\left( \bfhatb{r}_x \bfhatb{r}_y + \bfhatb{r}_y \bfhatb{r}_x \right) \left(F_x \bar{r}_y + F_y \bar{r}_x \right) \nonumber \\
		&\qquad \quad +\frac{1}{2}\left( \bfhatb{r}_x \bfhatb{r}_z + \bfhatb{r}_z \bfhatb{r}_x \right) \left(F_x \bar{r}_z + F_z \bar{r}_x \right) +\frac{1}{2}\left( \bfhatb{r}_y \bfhatb{r}_z + \bfhatb{r}_z \bfhatb{r}_y \right) \left(F_y \bar{r}_z + F_z \bar{r}_y \right) \Bigg] \nonumber \\
		& +\frac{\sqrt{5}}{42} Y_{00} \left[ (5-21) \left(F_x \bar{r}_x + F_y \bar{r}_y \right)+(11-21)\, F_z \bar{r}_z \right] \left( -1 + 3 \bfhatb{r}_z \bfhatb{r}_z \right) \label{eq:G4_simpl2}\\
		= & \frac{\sqrt{5}}{7} Y_{00} \Bigg[ -\left( \bfhatb{r}_x \bfhatb{r}_x + \bfhatb{r}_y \bfhatb{r}_y - \bfhatb{r}_z \bfhatb{r}_z\right) \left(F_x \bar{r}_x + F_y \bar{r}_y \right)  + \left( \bfhatb{r}_x \bfhatb{r}_x + \bfhatb{r}_y \bfhatb{r}_y \right) F_z \bar{r}_z \Bigg] \nonumber \\
		& -\frac{\sqrt{5}}{21} Y_{00} \left[ 8 \left(F_x \bar{r}_x + F_y \bar{r}_y \right)+5\, F_z \bar{r}_z \right] \left( -1 + 3 \bfhatb{r}_z \bfhatb{r}_z  \right) \nonumber\\
		= & \frac{\sqrt{5}}{7} Y_{00} \Bigg\{ -\left( 1 - 2\bfhatb{r}_z \bfhatb{r}_z\right) \left(F_x \bar{r}_x + F_y \bar{r}_y \right)  + \left( 1- \bfhatb{r}_z \bfhatb{r}_z \right) F_z \bar{r}_z  \nonumber \\
		& \qquad - \left[\frac{8}{3} \left(F_x \bar{r}_x + F_y \bar{r}_y \right)+\frac{5}{3}\, F_z \bar{r}_z \right] \left( -1 + 3 \bfhatb{r}_z \bfhatb{r}_z \right) \Bigg\} \nonumber\\
		= & \frac{\sqrt{5}}{7} Y_{00} \Bigg\{ \left( \frac{5}{3} - 6\bfhatb{r}_z \bfhatb{r}_z\right) \left(F_x \bar{r}_x + F_y \bar{r}_y \right)  + \left( \frac{8}{3}-6 \bfhatb{r}_z \bfhatb{r}_z \right) F_z \bar{r}_z \Bigg\} \nonumber\\
		= & \frac{\sqrt{5}}{7} Y_{00} \Bigg\{ \left( \frac{5}{3} - 6\bfhatb{r}_z \bfhatb{r}_z\right) \mbf{F} \cdot \mbfb{r}  + F_z \bar{r}_z \Bigg\}.
	\end{align}
	The new part of the term for $n \geq 4$ can also be simplified. First, we note the following identities that are derived in the same manner as those from Eq.~\eqref{eq:r_x}--\eqref{eq:rr_zz}:
	\begin{align}
		\bar{Y}_{42} + \bar{Y}_{4-2} &= \frac{3}{4} \sqrt{\frac{5}{2\pi}}\; \left(\bfhatb{r}_x \bfhatb{r}_x - \bfhatb{r}_y \bfhatb{r}_y \right) \left(-1+7\bfhatb{r}_z\bfhatb{r}_z\right), \\
		-i\left(\bar{Y}_{42} - \bar{Y}_{4-2} \right) &= \frac{3}{4} \sqrt{\frac{5}{2\pi}}\; \left(\bfhatb{r}_x \bfhatb{r}_y + \bfhatb{r}_y \bfhatb{r}_x \right) \left(-1+7\bfhatb{r}_z\bfhatb{r}_z\right), \\
		-\sqrt{2}\left(\bar{Y}_{41}- \bar{Y}_{4-1} \right) &= \frac{3}{4} \sqrt{\frac{5}{2\pi}}\; \left(\bfhatb{r}_x \bfhatb{r}_z + \bfhatb{r}_z \bfhatb{r}_x \right) \left(-3+7\bfhatb{r}_z\bfhatb{r}_z\right), \\
		\sqrt{2}i\left(\bar{Y}_{41} + \bar{Y}_{4-1} \right) &= \frac{3}{4} \sqrt{\frac{5}{2\pi}}\; \left(\bfhatb{r}_y \bfhatb{r}_z + \bfhatb{r}_z \bfhatb{r}_y \right) \left(-3+7\bfhatb{r}_z\bfhatb{r}_z\right), \\
		\bar{Y}_{40} &= \frac{3}{16\sqrt{\pi}} \left[35 \left( \bfhatb{r}_z \right)^4 - 30 \left( \bfhatb{r}_z \right)^2 +3 \right].
	\end{align}
	\newpage
	Then, the following simplifications are possible:
	{\allowdisplaybreaks
	\begin{align}
		&- \frac{1}{7\sqrt{2}} \Big[ \left( \bar{Y}_{42} + \bar{Y}_{4-2} \right) \left(F_x \bar{r}_x - F_y \bar{r}_y \right)
		-i \left( \bar{Y}_{42} - \bar{Y}_{4-2} \right) \left(F_x \bar{r}_y + F_y \bar{r}_x \right) \nonumber\\
		&\qquad -\sqrt{2} \left(\bar{Y}_{41} - \bar{Y}_{4-1}\right) \left(F_x \bar{r}_z + F_z \bar{r}_x \right) 
		+\sqrt{2}i \left(\bar{Y}_{41} + \bar{Y}_{4-1}\right) \left(F_y \bar{r}_z + F_z \bar{r}_y \right) \Big] \nonumber\\
		&- \frac{2}{7\sqrt{5}} \bar{Y}_{40} \Big[ -\left(F_x \bar{r}_x + F_y \bar{r}_y \right)+2\, F_z \bar{r}_z \Big] \nonumber\\
		= & \frac{3 \sqrt{5}}{28} Y_{00} \Bigg\{ \Big[ \left( \bfhatb{r}_x \bfhatb{r}_x - \bfhatb{r}_y \bfhatb{r}_y \right) \left(F_x \bar{r}_x - F_y \bar{r}_y \right)
		+ \left( \bfhatb{r}_x \bfhatb{r}_y + \bfhatb{r}_y \bfhatb{r}_x \right) \left(F_x \bar{r}_y + F_y \bar{r}_x \right)\Big] \nonumber\\
		& \qquad  \qquad \left(-1+7\bfhatb{r}_z\bfhatb{r}_z\right) + \Big[ \left( \bfhatb{r}_x \bfhatb{r}_z + \bfhatb{r}_z \bfhatb{r}_x \right) \left(F_x \bar{r}_z + F_z \bar{r}_x \right) \nonumber \\
		& \qquad \qquad + \left( \bfhatb{r}_y \bfhatb{r}_z + \bfhatb{r}_z \bfhatb{r}_y \right) \left(F_y \bar{r}_z + F_z \bar{r}_y \right) \Big] \left(-3+7\bfhatb{r}_z\bfhatb{r}_z\right) \Bigg\} \nonumber\\
		& + \frac{3 \sqrt{5}}{140} Y_{00} \left[35 \left( \bfhatb{r}_z \right)^4 - 30 \left( \bfhatb{r}_z \right)^2 +3 \right] \Big[ -\left(F_x \bar{r}_x + F_y \bar{r}_y \right)+2\, F_z \bar{r}_z \Big] \nonumber\\
		= & \frac{3 \sqrt{5}}{28} Y_{00} \Bigg\{ \left(F_x \bar{r}_x + F_y \bar{r}_y \right) \Bigg[ 7\left( \bfhatb{r}_x \bfhatb{r}_x + \bfhatb{r}_y \bfhatb{r}_y \right) \bfhatb{r}_z \bfhatb{r}_z - \left( \bfhatb{r}_x \bfhatb{r}_x + \bfhatb{r}_y \bfhatb{r}_y \right) \nonumber \\
		& \qquad \qquad \quad +14 \left( \bfhatb{r}_z \right)^4 - 6 \left( \bfhatb{r}_z \right)^2 -7 \left( \bfhatb{r}_z \right)^4 + 6 \left( \bfhatb{r}_z \right)^2-\frac{3}{5}\Bigg] \nonumber \\
		& \qquad \qquad + F_z \bar{r}_z \left[ 14\left( \bfhatb{r}_x \bfhatb{r}_x + \bfhatb{r}_y \bfhatb{r}_y \right) \bfhatb{r}_z \bfhatb{r}_z -6 \left( \bfhatb{r}_x \bfhatb{r}_x + \bfhatb{r}_y \bfhatb{r}_y \right) +14 \left( \bfhatb{r}_z \right)^4 - 12 \left( \bfhatb{r}_z \right)^2 +\frac{6}{5}\right] \Bigg\} \nonumber\\
		= & \frac{3 \sqrt{5}}{28} Y_{00} \Bigg\{ \left(F_x \bar{r}_x + F_y \bar{r}_y \right) \Bigg[ 7\left( 1 - \bfhatb{r}_z \bfhatb{r}_z \right) \bfhatb{r}_z \bfhatb{r}_z - \left( 1 - \bfhatb{r}_z \bfhatb{r}_z \right) +7 \left( \bfhatb{r}_z \right)^4 -\frac{3}{5}\Bigg] \nonumber \\
		& \qquad \qquad + F_z \bar{r}_z \left[ 14\left( 1 - \bfhatb{r}_z \bfhatb{r}_z \right) \bfhatb{r}_z \bfhatb{r}_z -6 \left( 1 - \bfhatb{r}_z \bfhatb{r}_z \right) +14 \left( \bfhatb{r}_z \right)^4 - 12 \left( \bfhatb{r}_z \right)^2 +\frac{6}{5}\right]\Bigg\} \nonumber\\
		= & \frac{3 \sqrt{5}}{28} Y_{00} \Bigg\{ \left(F_x \bar{r}_x + F_y \bar{r}_y \right) \left[ 8\bfhatb{r}_z \bfhatb{r}_z -\frac{8}{5}\right] + F_z \bar{r}_z \left[ 8\bfhatb{r}_z \bfhatb{r}_z -\frac{24}{5}\right]\Bigg\}. \label{eq:G4_simpl3}
	\end{align}
	}
	When we now sum the terms for $n=0$, the new terms  for $n=2$, and the new terms for $n \geq 4$ to get the total term for $n \geq 4$, we find:
	\begin{align}
		&\frac{-1}{3\sqrt{5}} Y_{00} \left[ -\left(F_x \bar{r}_x + F_y \bar{r}_y \right)+2 F_z \bar{r}_z \right] - \frac{\sqrt{5}}{7} Y_{00} \Bigg\{ \left( \frac{5}{3} - 6\bfhatb{r}_z \bfhatb{r}_z\right) \mbf{F} \cdot \mbfb{r}  + F_z \bar{r}_z \Bigg\} \nonumber\\
		& -\frac{3 \sqrt{5}}{28} Y_{00} \Bigg\{ 8\, \mbf{F} \cdot \mbfb{r} \; \bfhatb{r}_z \bfhatb{r}_z -\frac{8}{5} \left(F_x \bar{r}_x + F_y \bar{r}_y \right) -\frac{24}{5}F_z \bar{r}_z \Bigg\} \nonumber \\
		=& \frac{1}{\sqrt{5}} Y_{00} \Bigg\{  \left(F_x \bar{r}_x + F_y \bar{r}_y \right) \left(\frac{1}{3}-\frac{25}{21}+\frac{6}{7}\right) +F_z \bar{r}_z \left(\frac{-2}{3}-\frac{40}{21}+\frac{18}{7}\right)
		+\mbf{F} \cdot \mbfb{r} \; \bfhatb{r}_z \bfhatb{r}_z \left(\frac{30}{7}-\frac{30}{7}\right)\Bigg\} \nonumber\\
		=& 0.
	\end{align}
	Consequently, the terms from IV only contribute for $n=0$ and $n=2$ and not for $n \geq 4$.
	\newpage
	\item The terms here are quite similar to those in the previous case IV. Again, the term for $n=0$ is already simplified as far as possible. The new part of the term for $n=2$ (compared to $n=0$) for term a can be simplified analogously to Eq.~\eqref{eq:G4_simpl2} and proceeding in the same manner:
	\begin{align}
		& \frac{\sqrt{5}}{7} Y_{00} \Bigg[ -\left( \bfhatb{r}_x \bfhatb{r}_x - \bfhatb{r}_y \bfhatb{r}_y \right) \left(F_x \bar{r}_x - F_y \bar{r}_y \right)  -\left( \bfhatb{r}_x \bfhatb{r}_y + \bfhatb{r}_y \bfhatb{r}_x \right) \left(F_x \bar{r}_y + F_y \bar{r}_x \right) \nonumber \\
		&\qquad \quad +\frac{1}{2}\left( \bfhatb{r}_x \bfhatb{r}_z + \bfhatb{r}_z \bfhatb{r}_x \right) \left(F_x \bar{r}_z + F_z \bar{r}_x \right) +\frac{1}{2}\left( \bfhatb{r}_y \bfhatb{r}_z + \bfhatb{r}_z \bfhatb{r}_y \right) \left(F_y \bar{r}_z + F_z \bar{r}_y \right) \Bigg] \nonumber \\
		& +\frac{\sqrt{5}}{42} Y_{00} \left[ (5+21) \left(F_x \bar{r}_x + F_y \bar{r}_y \right)+(11+21)\, F_z \bar{r}_z \right] \left( -1 + 3 \bfhatb{r}_z \bfhatb{r}_z \right) \nonumber\\
		= & \frac{\sqrt{5}}{7} Y_{00} \Bigg\{ -\left( 1 - 2\bfhatb{r}_z \bfhatb{r}_z\right) \left(F_x \bar{r}_x + F_y \bar{r}_y \right)  + \left( 1- \bfhatb{r}_z \bfhatb{r}_z \right) F_z \bar{r}_z  \nonumber \\
		& \qquad + \left[\frac{13}{3} \left(F_x \bar{r}_x + F_y \bar{r}_y \right)+\frac{16}{3}\, F_z \bar{r}_z \right] \left( -1 + 3 \bfhatb{r}_z \bfhatb{r}_z \right) \Bigg\} \nonumber\\
		= & \frac{\sqrt{5}}{7} Y_{00} \Bigg\{ \left( -\frac{16}{3} +15\bfhatb{r}_z \bfhatb{r}_z\right) \left(F_x \bar{r}_x + F_y \bar{r}_y \right)  + \left( -\frac{13}{3} +15\bfhatb{r}_z \bfhatb{r}_z\right) F_z \bar{r}_z \Bigg\} \nonumber\\
		= & \frac{\sqrt{5}}{7} Y_{00} \Bigg\{ \left( -\frac{16}{3} +15\bfhatb{r}_z \bfhatb{r}_z\right) \mbf{F} \cdot \mbfb{r}  + F_z \bar{r}_z \Bigg\}.
	\end{align}
	For b, the new part of the term for $n=2$ is identical, but with a prefactor of $7$.\\
	The new part of the term for $n \geq 4$ can also be simplified --- it is exactly the same as in the previous case IV, but with a global negative sign. Therefore, we can use Eq.~\eqref{eq:G4_simpl3} again, adding a negative sign.
	When we now sum the terms for $n=0$, the new terms  for $n=2$, and the new terms for $n \geq 4$ to get the total term for $n \geq 4$ in a, we find:
	{\allowdisplaybreaks
	\begin{align}
		&\frac{1}{3\sqrt{5}} Y_{00} \left[ -\left(F_x \bar{r}_x + F_y \bar{r}_y \right)+2 F_z \bar{r}_z \right] + \frac{\sqrt{5}}{7} Y_{00} \Bigg\{ \left( -\frac{16}{3} +15\bfhatb{r}_z \bfhatb{r}_z\right) \mbf{F} \cdot \mbfb{r}  + F_z \bar{r}_z \Bigg\} \nonumber\\
		& +\frac{3 \sqrt{5}}{28} Y_{00} \Bigg\{ 8\, \mbf{F} \cdot \mbfb{r} \; \bfhatb{r}_z \bfhatb{r}_z -\frac{8}{5} \left(F_x \bar{r}_x + F_y \bar{r}_y \right) -\frac{24}{5}F_z \bar{r}_z \Bigg\} \nonumber \\
		= & \frac{1}{\sqrt{5}} Y_{00} \left\{  \left(F_x \bar{r}_x + F_y \bar{r}_y \right) \left(-\frac{1}{3}-\frac{80}{21}-\frac{6}{7}\right) +F_z \bar{r}_z \left(\frac{2}{3}-\frac{65}{21}-\frac{18}{7}\right) 
		+\mbf{F} \cdot \mbfb{r} \; \bfhatb{r}_z \bfhatb{r}_z \left(\frac{75}{7}+\frac{30}{7}\right)\right\} \nonumber\\
		=& \frac{1}{\sqrt{5}} Y_{00} \left\{  -5 \left(F_x \bar{r}_x + F_y \bar{r}_y \right)  -5 F_z \bar{r}_z  +15\, \mbf{F} \cdot \mbfb{r} \; \bfhatb{r}_z \bfhatb{r}_z \right\} \nonumber\\
		=& \sqrt{5} \; Y_{00} \, \mbf{F} \cdot \mbfb{r} \left(-1+3\bfhatb{r}_z\bfhatb{r}_z\right).
	\end{align}
	}
	For b, we express the new terms for $n \geq 4$ as this sum minus the contributions from $n=0$ and $n=2$.\\
	\end{enumerate}
	\newpage
	Finally, we summarize all these terms into one equation. The first curly brackets encompass the contributions from term I and II, the second one from term III, the third one from term IV, and the fourth one from term V. The following (lengthy) equation results for $u^{\bot}_{20}$:
	{\allowdisplaybreaks
		\begin{alignat}{2}
			\frac{16\pi\mu(1-\nu)}{4\pi \; Y_{00}} R^3 u^{\bot}_{20}=& \sqrt{\frac{1}{5}} R^2 \Bigg\{ \!\!&&\left( \frac{1}{3} k_1 -4 \alpha_1 + 5h_1\right) \left(-F_x \bfhatb{r}_x - F_y \bfhatb{r}_y + 2 F_z \bfhatb{r}_z\right) \frac{\bar{r}}{R}\nonumber \\
			& &&+\frac{1}{7} k_3 \frac{3}{2} \left(-F_x \bfhatb{r}_x - F_y \bfhatb{r}_y - 3 F_z \bfhatb{r}_z +5 \bfhatb{r}_z \bfhatb{r}_z \, \mbf{F} \cdot \bfhatb{r} \right) \left(\frac{\bar{r}}{R}\right)^3 \nonumber\\
			& && +\sum_{\substack{n=3\\n \text{ odd}}}^{\infty} \Bigg[ \left( \alpha_{n} +(2n+3) h_{n} -\frac{(n+1)(n+4)}{2} \alpha_{n} \right) \nonumber \\
			& && \qquad \qquad \left(-F_x \bfhatb{r}_x - F_y \bfhatb{r}_y + 2 F_z \bfhatb{r}_z\right) \nonumber \\
			& && \qquad \qquad + \left( \alpha_{n} +(2n+3) h_{n} -\frac{(n-1)(n+6)}{2} \alpha_{n} \right) \nonumber \\
			& && \qquad \qquad \frac{3}{2} \left(-F_x \bfhatb{r}_x - F_y \bfhatb{r}_y - 3 F_z \bfhatb{r}_z +5 \bfhatb{r}_z \bfhatb{r}_z \, \mbf{F} \cdot \bfhatb{r} \right) \Bigg] \left(\frac{\bar{r}}{R}\right)^n\Bigg\} \nonumber\\
			& + \sqrt{\frac{1}{5}} \bar{r} && \left( -1 + 3 \bfhatb{r}_z \bfhatb{r}_z \right) \mbf{F} \cdot \mbfb{r} \Bigg\{ \left( m_3 - 5 \alpha_3 \right) \frac{\bar{r}}{R} \nonumber\\
			& && + \sum_{\substack{n=3\\n \text{ odd}}}^{\infty} \Bigg[ \frac{5}{2}m_{n+2} - \alpha_{n+2} \Bigg(\frac{(n+1)(n+4)}{2} + \frac{3}{2} \frac{(n-1)(n+6)}{2} \Bigg) \Bigg] \left(\frac{\bar{r}}{R}\right)^n\Bigg\} \nonumber \\
			& - R \Bigg\{&& \left[\beta_0 + \beta_2 \left(\frac{\bar{r}}{R}\right)^2\right]  \frac{\bar{r}}{3\sqrt{5}} \left(-F_x \bfhatb{r}_x - F_y \bfhatb{r}_y + 2 F_z \bfhatb{r}_z\right) \nonumber\\
			& && + \beta_2 \frac{\sqrt{5}}{7} \Bigg[ \left( \frac{5}{3} - 6\bfhatb{r}_z \bfhatb{r}_z\right) \mbf{F} \cdot \mbfb{r}  + F_z \bar{r}_z \Bigg] \left(\frac{\bar{r}}{R}\right)^2\Bigg\} \nonumber \\ 
			&+ R \Bigg\{&& \left[-\gamma_0 - \gamma_2 \left(\frac{\bar{r}}{R}\right)^2 + 3 \alpha_1 + 10 \alpha_3 \left(\frac{\bar{r}}{R}\right)^2\right] \nonumber\\
			& && \frac{\bar{r}}{3\sqrt{5}} \left(-F_x \bfhatb{r}_x - F_y \bfhatb{r}_y + 2 F_z \bfhatb{r}_z\right) 
			+ \left( -\gamma_2 + 7 \alpha_3 \right) \left(\frac{\bar{r}}{R}\right)^2 \nonumber\\
			& && \frac{\sqrt{5}}{7} \Bigg[ \left( -\frac{16}{3} +15\bfhatb{r}_z \bfhatb{r}_z\right) \mbf{F} \cdot \mbfb{r}  + F_z \bar{r}_z \Bigg] \nonumber\\
			& && +\sum_{\substack{n=4\\n \text{ even}}}^{\infty} \Bigg[ \left( -\gamma_n + \alpha_{n+1} \frac{(n-2)(n+7)}{2} \right)
			\sqrt{5} \; \mbf{F} \cdot \mbfb{r} \left(-1+3\bfhatb{r}_z\bfhatb{r}_z\right) \nonumber\\
			& &&\qquad \qquad + \alpha_{n+1} \left( \frac{(n+2)(n+3)}{2} -\frac{(n-2)(n+7)}{2} \right) \nonumber\\
			& &&\qquad \qquad \frac{\bar{r}}{3\sqrt{5}} \left(-F_x \bfhatb{r}_x - F_y \bfhatb{r}_y + 2 F_z \bfhatb{r}_z\right) \nonumber\\
			& &&\qquad \qquad + \alpha_{n+1} \left( \frac{n(n+5)}{2} -\frac{(n-2)(n+7)}{2} \right) \nonumber\\
			& &&\qquad \qquad \frac{\sqrt{5}}{7} \left[ \left( -\frac{16}{3} +15\bfhatb{r}_z \bfhatb{r}_z\right) \mbf{F} \cdot \mbfb{r}  + F_z \bar{r}_z \right] \Bigg] \left(\frac{\bar{r}}{R}\right)^n\Bigg\}. \label{eq:u20_1}
		\end{alignat}
	}
	To simplify these expressions, we rewrite some of the expressions containing polynomials in $n$:
	\begin{align}
		\frac{(n+1)(n+4)}{2} + \frac{3}{2} \frac{(n-1)(n+6)}{2} &= \frac{5}{4} \left( (n+2)^2+(n+2)-8 \right), \\
		\frac{(n+2)(n+3)}{2} -\frac{(n-2)(n+7)}{2} &= 10, \\
		\frac{n(n+5)}{2} -\frac{(n-2)(n+7)}{2} &= 7.
	\end{align}
	Additionally, we summarize the terms with prefactor $\alpha_{n} +(2n+3) h_{n}$ in the first curly brackets:
	\begin{align}
		& \left(-F_x \bfhatb{r}_x - F_y \bfhatb{r}_y + 2 F_z \bfhatb{r}_z\right) + \frac{3}{2} \left(-F_x \bfhatb{r}_x - F_y \bfhatb{r}_y - 3 F_z \bfhatb{r}_z +5 \bfhatb{r}_z \bfhatb{r}_z \, \mbf{F} \cdot \bfhatb{r} \right) \nonumber\\
		= & \frac{5}{2} \mbf{F} \cdot \bfhatb{r} \left(-1+3\bfhatb{r}_z\bfhatb{r}_z\right).
	\end{align}
	Furthermore, we notice the following common factors in some terms that we factor out:
	\begin{align}
		& -F_x \bfhatb{r}_x - F_y \bfhatb{r}_y + 2 F_z \bfhatb{r}_z, \\
		& -F_x \bfhatb{r}_x - F_y \bfhatb{r}_y - 3 F_z \bfhatb{r}_z +5 \bfhatb{r}_z \bfhatb{r}_z \, \mbf{F} \cdot \bfhatb{r}, \\
		& \mbf{F} \cdot \mbfb{r} \left(-1+3\bfhatb{r}_z\bfhatb{r}_z\right).
	\end{align}
	Then, we can simplify Eq.~\eqref{eq:u20_1} as
	{\allowdisplaybreaks
	\begin{align}
		\frac{4\mu(1-\nu)}{Y_{00}} \sqrt{5} R^3 u^{\bot}_{20}=& \left(-F_x \bfhatb{r}_x - F_y \bfhatb{r}_y + 2 F_z \bfhatb{r}_z\right) \Bigg\{ \!\!\left( \frac{1}{3} k_1 -4 \alpha_1 + 5h_1\right)  R \bar{r} \nonumber \\
		& \qquad - \left[\beta_0 + \beta_2 \left(\frac{\bar{r}}{R}\right)^2\right] \frac{R \bar{r}}{3} 
		+ \left[-\gamma_0 - \gamma_2 \left(\frac{\bar{r}}{R}\right)^2 + 3 \alpha_1 \right] \frac{R \bar{r}}{3}\nonumber\\
		& \qquad +\sum_{\substack{n=3\\n \text{ odd}}}^{\infty} \left(-\frac{(n+1)(n+4)}{2} \alpha_{n} + \frac{10}{3} \alpha_{n} \right) R^2 \left(\frac{\bar{r}}{R}\right)^n\Bigg\} \nonumber \\
		& +\left(-F_x \bfhatb{r}_x - F_y \bfhatb{r}_y - 3 F_z \bfhatb{r}_z +5 \bfhatb{r}_z \bfhatb{r}_z \, \mbf{F} \cdot \bfhatb{r} \right) \frac{3}{2}R^2 \nonumber\\
		& \qquad \left\{ \frac{1}{7} k_3 \left(\frac{\bar{r}}{R}\right)^3 -\sum_{\substack{n=3\\n \text{ odd}}}^{\infty} \left(\frac{(n-1)(n+6)}{2} \alpha_{n} \right) \left(\frac{\bar{r}}{R}\right)^n \right\} \nonumber \\
		& + \left( -1 + 3 \bfhatb{r}_z \bfhatb{r}_z \right) \mbf{F} \cdot \mbfb{r} \Bigg\{ \left( m_3 - 5 \alpha_3 \right) \frac{\bar{r}}{R} \, \bar{r} \nonumber\\
		&  \qquad + \frac{5}{2} \sum_{\substack{n=3\\n \text{ odd}}}^{\infty} \left[ \alpha_{n} +(2n+3) h_{n} \right] \left(\frac{\bar{r}}{R}\right)^{n-1} R \nonumber\\
		&  \qquad + \frac{5}{2} \sum_{\substack{n=5\\n \text{ odd}}}^{\infty} \Bigg[ m_{n} - \alpha_{n} \frac{n^2+n-8}{2} \Bigg] \left(\frac{\bar{r}}{R}\right)^{n-2} \bar{r} \nonumber\\
		&  \qquad + 5 \sum_{\substack{n=5\\n \text{ odd}}}^{\infty} \Bigg[ -\gamma_{n-1} +\frac{(n-3)(n+6)}{2} \alpha_{n} \Bigg] \left(\frac{\bar{r}}{R}\right)^{n-1} R \Bigg\} \nonumber \\
		& - R \, \beta_2 \left(\frac{\bar{r}}{R}\right)^2 \frac{5}{7} \Bigg[ \left( \frac{5}{3} - 6\bfhatb{r}_z \bfhatb{r}_z\right) \mbf{F} \cdot \mbfb{r}  + F_z \bar{r}_z \Bigg] \nonumber \displaybreak\\ 
		&+ \frac{5}{7} R \Bigg[ \left( -\frac{16}{3} +15\bfhatb{r}_z \bfhatb{r}_z\right) \mbf{F} \cdot \mbfb{r}  + F_z \bar{r}_z \Bigg] \nonumber\\
		& \qquad \Bigg\{  -\gamma_2 \left(\frac{\bar{r}}{R}\right)^2 + 7 \sum_{\substack{n=3\\n \text{ odd}}}^{\infty} \alpha_{n} \left(\frac{\bar{r}}{R}\right)^{n-1}\Bigg\}. \label{eq:u20_2}
	\end{align}
	}
	Next, we consider the terms in the series expansions for $n \geq 5$ separately. If all of those terms vanish when summed up, we would only need a finite number of terms to calculate $u^{\bot}_{20}$. In this calculation, we recall Eq.~\eqref{eq:gamma_simpl} so that three terms cancel each other [those with prefactor $\gamma_{n-1}$, $m_n$ and $(2n+3)h_n$]. Furthermore, we note that $\left(\frac{\bar{r}}{R}\right)^{n} R^2 = \left(\frac{\bar{r}}{R}\right)^{n-1} R \bar{r} = \left(\frac{\bar{r}}{R}\right)^{n-2} (\bar{r})^2$ and $\bfhatb{r} = \mbfb{r} / \bar{r}$, so that we can factor out this common term. The remaining part on the right-hand side [without sum and factor $\left(\frac{\bar{r}}{R}\right)^{n} R^2$] is then
	\begin{align}
		& \left(-F_x \bfhatb{r}_x - F_y \bfhatb{r}_y + 2 F_z \bfhatb{r}_z\right) \left[-\frac{(n+1)(n+4)}{2} \alpha_{n} + \frac{10}{3} \alpha_{n} \right] \nonumber\\
		& -\left(-F_x \bfhatb{r}_x - F_y \bfhatb{r}_y - 3 F_z \bfhatb{r}_z +5 \bfhatb{r}_z \bfhatb{r}_z \, \mbf{F} \cdot \bfhatb{r} \right) \frac{3}{2} \left[\frac{(n-1)(n+6)}{2} \alpha_{n} \right] \nonumber\\
		& + \frac{5}{2} \left( -1 + 3 \bfhatb{r}_z \bfhatb{r}_z \right) \mbf{F} \cdot \bfhatb{r} \left[ \alpha_{n} - \alpha_{n} \frac{n^2+n-8}{2} + (n-3)(n+6) \alpha_{n} \right] \nonumber\\
		& + \frac{5}{7} R \left[ \left( -\frac{16}{3} +15\bfhatb{r}_z \bfhatb{r}_z\right) \mbf{F} \cdot \bfhatb{r}  + F_z \bfhatb{r}_z \right] 7\alpha_{n}  \nonumber\\
		= & \mbf{F} \cdot \bfhatb{r} \; \alpha_{n} \Bigg[\frac{(n+1)(n+4)}{2} - \frac{10}{3} + \frac{3}{2} \frac{(n-1)(n+6)}{2} -\frac{5}{2} + \frac{5}{2} \frac{n^2+n-8}{2} 
		- \frac{5}{2} (n-3)(n+6) - \frac{80}{3}\Bigg]  \nonumber\\
		& + F_z \bfhatb{r}_z \; \alpha_{n} \Bigg[-3\frac{(n+1)(n+4)}{2} +10 + 3 \frac{(n-1)(n+6)}{2} +5 \Bigg]  \nonumber\\
		& + \bfhatb{r}_z \bfhatb{r}_z \, \mbf{F} \cdot \bfhatb{r} \; \alpha_{n} \Bigg[ -\frac{15}{2} \frac{(n-1)(n+6)}{2} + \frac{15}{2} - \frac{15}{2} \frac{n^2+n-8}{2} 
		+ \frac{15}{2} (n-3)(n+6) + 75 \Bigg]  \nonumber\\
		= & 0.
	\end{align}
	In the last equation, expanding all the square brackets shows that these three polynomials in $n$ all sum to zero.
	\newpage
	Therefore, only a finite number of terms contribute to Eq.~\eqref{eq:u20_2}. We simplify these terms as fellows, analogously to the calculation for the terms for $n \geq 5$:
	{\allowdisplaybreaks
		\begin{align}
			\frac{4\mu(1-\nu)}{Y_{00}} \sqrt{5} R^3 u^{\bot}_{20}=& \left(-F_x \bfhatb{r}_x - F_y \bfhatb{r}_y + 2 F_z \bfhatb{r}_z\right) \Bigg\{ \!\!\left( \frac{1}{3} k_1 -4 \alpha_1 + 5h_1\right)  R \bar{r} \nonumber \\
			& \qquad - \left[\beta_0 + \beta_2 \left(\frac{\bar{r}}{R}\right)^2\right] \frac{R \bar{r}}{3}
			+ \left[-\gamma_0 - \gamma_2 \left(\frac{\bar{r}}{R}\right)^2 + 3 \alpha_1 \right] \frac{R \bar{r}}{3}\nonumber\\
			& \qquad - \frac{32}{3} \alpha_{3} R^2 \left(\frac{\bar{r}}{R}\right)^3\Bigg\} \nonumber \\
			& +\left(-F_x \bfhatb{r}_x - F_y \bfhatb{r}_y - 3 F_z \bfhatb{r}_z +5 \bfhatb{r}_z \bfhatb{r}_z \, \mbf{F} \cdot \bfhatb{r} \right) \frac{3}{2}R^2 \nonumber\\
			& \qquad \left\{ \frac{1}{7} k_3 \left(\frac{\bar{r}}{R}\right)^3 -9 \alpha_{3} \left(\frac{\bar{r}}{R}\right)^3 \right\} \nonumber \\
			& + \left( -1 + 3 \bfhatb{r}_z \bfhatb{r}_z \right) \mbf{F} \cdot \mbfb{r} \Bigg\{ \left( m_3 - 5 \alpha_3 \right) \frac{\bar{r}}{R} \, \bar{r} 
			+ \frac{5}{2} \left[ \alpha_{3} + 9 h_{3} \right] \left(\frac{\bar{r}}{R}\right)^{2} R  \Bigg\} \nonumber \\
			& - R \, \beta_2 \left(\frac{\bar{r}}{R}\right)^2 \frac{5}{7} \left[ \left( \frac{5}{3} - 6\bfhatb{r}_z \bfhatb{r}_z\right) \mbf{F} \cdot \mbfb{r}  + F_z \bar{r}_z \right] \nonumber \\ 
			&+ \frac{5}{7} R \left[ \left( -\frac{16}{3} +15\bfhatb{r}_z \bfhatb{r}_z\right) \mbf{F} \cdot \mbfb{r}  + F_z \bar{r}_z \right] 
			\Bigg\{  -\gamma_2 \left(\frac{\bar{r}}{R}\right)^2 + 7 \alpha_{3} \left(\frac{\bar{r}}{R}\right)^{2}\Bigg\} \nonumber \\ 
			= & \left(-F_x \bfhatb{r}_x - F_y \bfhatb{r}_y + 2 F_z \bfhatb{r}_z\right) R \bar{r}\left[ \frac{1}{3} k_1 - 3 \alpha_1 + 5h_1 - \frac{1}{3} (\beta_0 +\gamma_0)\right]   \nonumber \\
			& + \mbf{F} \cdot \bfhatb{r} \, \frac{\bar{r}^3}{R} \Bigg[ \frac{1}{3} \beta_2 -\frac{25}{21} \beta_2 + \frac{1}{3} \gamma_2 + \frac{80}{21} \gamma_2 - \frac{3}{14} k_3 -m_3 - \frac{45}{2} h_3 \nonumber \\
			& \qquad \qquad + \alpha_{3} \left( \frac{32}{3}+\frac{27}{2}+5-\frac{5}{2}-\frac{80}{3}\right)\Bigg] \nonumber \\
			& + F_z \bfhatb{r}_z \, \frac{\bar{r}^3}{R} \left[ -\beta_2 -\frac{5}{7} \beta_2 -\gamma_2 -\frac{5}{7} \gamma_2 -\frac{3}{7} k_3 + \alpha_{3} \left(-32+27+5\right)\right] \nonumber \\
			& + \bfhatb{r}_z \bfhatb{r}_z \, \mbf{F} \cdot \bfhatb{r} \, \frac{\bar{r}^3}{R} \Bigg[ \frac{15}{14} k_3 + 3 m_3 +\frac{135}{2} h_3 +\frac{30}{7} \beta_2 -\frac{75}{7} \gamma_2 \nonumber \\
			& \qquad \qquad + \alpha_{3} \left(-\frac{135}{2} -15+\frac{15}{2}+75\right) \Bigg] \nonumber\\
			= & \left(-F_x \bfhatb{r}_x - F_y \bfhatb{r}_y + 2 F_z \bfhatb{r}_z\right) R \bar{r}
			\left[ \frac{1}{3} k_1 - 3 \alpha_1 + 5h_1 - \frac{1}{3} (\beta_0 +\gamma_0)\right]   \nonumber \\
			& + \mbf{F} \cdot \bfhatb{r} \, \frac{\bar{r}^3}{R} \left[ -\frac{6}{7} \beta_2 + \frac{29}{7} \gamma_2 - \frac{3}{14} k_3 -m_3 - \frac{45}{2} h_3 \right] \nonumber \\
			& + F_z \bfhatb{r}_z \, \frac{\bar{r}^3}{R} \left[ -\frac{12}{7} \left(\beta_2+\gamma_2 \right) -\frac{3}{7} k_3 \right]\nonumber \\
			& + \bfhatb{r}_z \bfhatb{r}_z \, \mbf{F} \cdot \bfhatb{r} \, \frac{\bar{r}^3}{R} \left[ \frac{15}{14} k_3 + 3 m_3 +\frac{135}{2} h_3 +\frac{30}{7} \beta_2 -\frac{75}{7} \gamma_2\right] \displaybreak\nonumber\\
			= & \left(-F_x \bfhatb{r}_x - F_y \bfhatb{r}_y + 2 F_z \bfhatb{r}_z\right) R \bar{r} \nonumber \\
			& \left[ \frac{1}{3} k_1 - 3 \left( h_1 - e_1 - u_3 \right) + 5h_1 - \frac{1}{6} \left( l_1 - 5h_1 + 5h_1 + m_1 \right) \right]   \nonumber \\
			& + \mbf{F} \cdot \bfhatb{r} \, \frac{\bar{r}^3}{R} \left[ -\frac{3}{7} \left( l_3 - 9h_3\right) + \frac{29}{14} \left( 9h_3 + m_3 \right) - \frac{3}{14} k_3 -m_3 - \frac{45}{2} h_3 \right] \nonumber \\
			& + F_z \bfhatb{r}_z \, \frac{\bar{r}^3}{R} \left[ -\frac{6}{7} \left( l_3 - 9h_3 + 9h_3 + m_3 \right) -\frac{3}{7} k_3 \right] \nonumber \\
			& + \bfhatb{r}_z \bfhatb{r}_z \, \mbf{F} \cdot \bfhatb{r} \, \frac{\bar{r}^3}{R} \left[ \frac{15}{14} k_3 + 3 m_3 +\frac{135}{2} h_3 +\frac{15}{7} \left( l_3 - 9h_3\right) -\frac{75}{14} \left( 9h_3 + m_3 \right) \right] \nonumber\\
			= & \left(-F_x \bfhatb{r}_x - F_y \bfhatb{r}_y + 2 F_z \bfhatb{r}_z\right) R \bar{r} \left[ \frac{1}{3} k_1 + 2h_1 + 3 e_1 + 3 u_3 - \frac{1}{6} l_1 - \frac{1}{6} m_1 \right]   \nonumber \\
			& + \left( F_x \bfhatb{r}_x + F_y \bfhatb{r}_y \right) \, \frac{\bar{r}^3}{R} \left[ -\frac{3}{7} l_3 + \frac{15}{14} m_3 - \frac{3}{14} k_3 \right] 
			+ F_z \bfhatb{r}_z \, \frac{\bar{r}^3}{R} \left[ -\frac{9}{7} l_3 + \frac{3}{14} m_3 - \frac{9}{14} k_3 \right] \nonumber \\
			& + \bfhatb{r}_z \bfhatb{r}_z \, \mbf{F} \cdot \bfhatb{r} \, \frac{\bar{r}^3}{R} \left[ \frac{15}{7} l_3 - \frac{33}{14} m_3 + \frac{15}{14} k_3 \right]. \label{eq:u20_4}
		\end{align}
	}
	In the second-to-last equation, we have used Eqs.~\eqref{eq:beta_0} and \eqref{eq:gamma_0} from before as well as (from the definitions)
	\begin{align}
		\alpha_1 &= h_1 - e_1 - u_3, \label{eq:alpha_1} \\
		\beta_2 &= \frac{1}{2} \left[ l_{3}-9h_{3} \right], \label{eq:beta_2}\\
		\gamma_2 &= \frac{1}{2} \left[ 9h_{3}+m_{3} \right]. \label{eq:gamma_2}
	\end{align}
	Finally, we insert the definitions from Eqs.~\eqref{eq:constants_1}--\eqref{eq:constants_14} and simplify with the aid of Mathematica \cite{Mathematica}:
	\begin{align}
		\frac{4\mu(1-\nu)}{Y_{00}} \sqrt{5} R^3 u^{\bot}_{20}= & \left(-F_x \bfhatb{r}_x - F_y \bfhatb{r}_y + 2 F_z \bfhatb{r}_z\right) R \bar{r} \nonumber \\
		& \frac{5 \mu (1-\nu)}{\mu\left( 7+5\nu\right) +4\tilde{\mu}\left( 7-10\nu\right) } \frac{2\mu\left( 7+2\nu\right)\left(4-5\tilde{\nu} \right) + 7\tilde{\mu}\left(17-19\tilde{\nu}\right) - 32\tilde{\mu}\nu\left(4-5\tilde{\nu}\right)}
		{2\mu\left(4-5\tilde{\nu} \right) + \tilde{\mu}(7-5\tilde{\nu})}  \nonumber \\[.5\baselineskip]
		& + \left( F_x \bfhatb{r}_x + F_y \bfhatb{r}_y \right) \, \frac{\bar{r}^3}{R} \, \frac{30 \mu (1-\nu) \nu}{\mu\left( 7+5\nu\right) +4\tilde{\mu}\left( 7-10\nu\right) } \nonumber \\[.5\baselineskip]
		& + F_z \bfhatb{r}_z \, \frac{\bar{r}^3}{R} \, \frac{-15 \mu (1-\nu) (7-6\nu)}{\mu\left( 7+5\nu\right) +4\tilde{\mu}\left( 7-10\nu\right) } \nonumber \\[.5\baselineskip]
		& + \bfhatb{r}_z \bfhatb{r}_z \, \mbf{F} \cdot \bfhatb{r} \, \frac{\bar{r}^3}{R} \, \frac{15 \mu (1-\nu) (7-10\nu)}{\mu\left( 7+5\nu\right) +4\tilde{\mu}\left( 7-10\nu\right) }. \label{eq:u20_5}
	\end{align}
	\newpage
	We notice that the solution is not isotropic anymore, but the dependence on $\left( F_x \bfhatb{r}_x + F_y \bfhatb{r}_y \right)$ is different from the dependence on $F_z \bfhatb{r}_z$. This is not surprising because the spherical harmonic $Y_{20}$ that we used in the expansion for this mode is symmetric under exchange of $x$- and $y$-components, but the $z$-component is different. The purpose of the deformational mode $u^{\bot}_{20}$ is precisely to measure the elongation along the $z$-axis relative to lateral contraction.\\
	As a last step of simplification, we write out $Y_{00}$ and introduce $ \underline{\bsy{\delta}}^{\parallel} = \bfhat{z} \bfhat{z}$, $ \underline{\bsy{\delta}}^{\perp} = \mathbb{1} - \underline{\bsy{\delta}}^{\parallel} $, where $ \bfhat{z} \bfhat{z}$ is a dyadic product, and $\mathbb{1}$ the unit matrix. Then, we obtain
	\begin{align}
		u^{\bot}_{20} &= \frac{\sqrt{5/4\pi}}{4 R^2} \frac{1}{\mu\left( 7+5\nu\right) +4\tilde{\mu}\left( 7-10\nu\right) } \nonumber\\
		&\hspace*{0.5cm} \Big[(-\mbf{F} \cdot \underline{\bsy{\delta}}^{\perp} \cdot \mbfb{r} + 2\, \mbf{F} \cdot \underline{\bsy{\delta}}^{\parallel} \cdot \mbfb{r}) 
		\frac{2\mu\left( 7+2\nu\right)\left(4-5\tilde{\nu} \right) + 7\tilde{\mu}\left(17-19\tilde{\nu}\right) - 32\tilde{\mu}\nu\left(4-5\tilde{\nu}\right)}
		{2\mu\left(4-5\tilde{\nu} \right) + \tilde{\mu}(7-5\tilde{\nu})} \nonumber\\
		&\hspace*{0.7cm} + 3\frac{\bar{r}^3}{R^2} \Big( 2\nu\, \mbf{F} \cdot \underline{\bsy{\delta}}^{\perp} \cdot \bfhatb{r} -(7-6\nu) \mbf{F} \cdot \underline{\bsy{\delta}}^{\parallel} \cdot \bfhatb{r} + \mbf{F} \cdot \bfhatb{r}\, \bfhatb{r} \cdot \underline{\bsy{\delta}}^{\parallel} \cdot \bfhatb{r} (7-10\nu)\Big) \Big]. \label{eq:u20_full}
	\end{align}
	In the limit of a free-standing sphere, $ \tilde{\mu} \rightarrow 0 $ (and for any value of $ \tilde{\nu} $), we obtain
	\begin{alignat}{3}
		u^{\bot}_{20} &= \frac{\sqrt{5/4\pi}}{4\mu R^2 (7+5\nu)} \Big[&&(-\mbf{F} \cdot \underline{\bsy{\delta}}^{\perp} \cdot \mbfb{r} + 2\, \mbf{F} \cdot \underline{\bsy{\delta}}^{\parallel} \cdot \mbfb{r}) \left( 7+2\nu\right) \nonumber\\
		& &&+ 3\frac{\bar{r}^3}{R^2} \left( 2\nu\, \mbf{F} \cdot \underline{\bsy{\delta}}^{\perp} \cdot \bfhatb{r} -(7-6\nu) \mbf{F} \cdot \underline{\bsy{\delta}}^{\parallel} \cdot \bfhatb{r} 
		+ \mbf{F} \cdot \bfhatb{r}\, \bfhatb{r} \cdot \underline{\bsy{\delta}}^{\parallel} \cdot \bfhatb{r} (7-10\nu)\right) \Big]. \label{eq:u20_manuscript}
	\end{alignat}
	As in the case of $u^{\bot}_{00}$, we have again confirmed that first taking the limit of $\tilde{\mu} \rightarrow 0$ (obtaining the solution for a free-standing sphere \cite{fischer2019magnetostriction}) and afterwards performing the same calculation as above leads to the same result as in Eq.~\eqref{eq:u20_manuscript}.
	
\section{Conclusions}
\label{sec:Conclusions}
In summary, we derive explicit analytical expressions that quantify the overall change in volume and the relative uniaxial elongation or contraction of a linearly elastic, isotropic, homogeneous sphere subject to the action of a discrete set of internal force centers. In that sense, our expressions bridge the scale from a potentially micro- or mesoscopic internal application of forces to the possibly macroscopic scale of deformation. Both cases of a free-standing elastic sphere and an elastic sphere embedded under no-slip conditions in an infinitely extended, homogeneous, isotropic, linearly elastic background are considered. 

Technically, our quantification works by calculating the amplitudes of spherical harmonics that quantify the normal displacements on the surface of the sphere. Our expressions provide a major simplification in determining these coefficients. Instead of first evaluating the displacement fields on various surface points and fitting the spherical harmonics \cite{fischer2019magnetostriction}, we now are able to calculate the coefficients directly from the internal force distribution. 

These relations directly motivate possible routes of future expansion of our work. Naturally, the derivation of expressions for the amplitudes of other modes of surface displacements should be explored as well, for instance, corresponding to twist deformational modes \cite{fischer2020} or higher, more strongly modulated modes of surface displacement \cite{fischer2024}. Besides, our analytical expressions are similar in spirit to Green’s functions that contain the information of the entire linear response of a system to a localized internal stimulus. Therefore, instead of summing over the effect of discrete individual force centers, we may by superposition likewise consider a continuous field of force distribution. Its effect is then included by integration. 

A first step of application of our expressions concerns problems of optimization. We wish to determine configurations of internal force centers that lead to maximized amplitudes of deformational response, in our case, change in volume and/or overall relative elongation or contraction along a given axis \cite{fischer2024misc}. In that case, we determine the extrema of our analytical expressions as a function of the positioning of the discrete internal force centers. Such strategies are important in the context of computation material optimization, here specifically in the field of stimuli-responsive soft actuators \cite{trivedi2008soft, whitesides2018soft, kim2019review}. 

\begin{acknowledgments}
	We thank the German Research Foundation (Deutsche Forschungsgemeinschaft, DFG) for support through the Heisenberg Grant no.\ ME 3571/4-1.
\end{acknowledgments}
\bibliography{literature.bib}
\end{document}